\documentclass[sigconf]{acmart}

\AtBeginDocument{
  }

\copyrightyear{2026}
\acmYear{2026}
\setcopyright{cc}
\setcctype{by}
\acmConference[UIST '26]{The 39th Annual ACM Symposium on User Interface Software and Technology}{November 02--05, 2026}{Detroit, MI, USA}
\acmBooktitle{The 39th Annual ACM Symposium on User Interface Software and Technology (UIST '26), November 02--05, 2026, Detroit, MI, USA}
\acmDOI{10.1145/3830398.3830631}
\acmISBN{979-8-4007-2856-3/2026/11}

\usepackage{enumitem}
\usepackage{subcaption}
\usepackage{multirow}
\usepackage{tikz}
\usepackage{xcolor}
\usepackage[normalem]{ulem}
\usepackage{listings}

\definecolor{orange_highlight}{RGB}{255,214,165}
\definecolor{purple_highlight}{RGB}{208,191,255}
\definecolor{incremental_diff}{RGB}{184,137,46}
\definecolor{insertion_diff}{RGB}{0,128,0}
\definecolor{deletion_diff}{RGB}{211,47,47}
\definecolor{primary_button}{RGB}{0,122,204}
\definecolor{secondary_button}{RGB}{95,106,121}
\definecolor{function_keyword}{RGB}{121,94,38}
\definecolor{mouse_selection}{RGB}{189,215,251}

\lstdefinestyle{prompt}{
  basicstyle=\ttfamily\small,
  lineskip=-1pt,
  breaklines=true,
  columns=fullflexible,
  literate={•}{$\bullet$}1
           {◦}{$\circ$}1
}

\DeclareRobustCommand{\captionTitle}[1]{%
  \tikz[baseline=(X.base)]{\node[rounded corners=5pt, fill=black!10,
    inner xsep=3pt, inner ysep=0pt] (X) {\strut\bfseries\sffamily #1};}%
}

\DeclareRobustCommand{\captionBox}[1]{%
  \tikz[baseline=(X.base)]{\node[rounded corners=5pt, fill=white,
    draw=black!60, line width=0.5pt, inner xsep=3pt, inner ysep=-0.5pt]
    (X) {\strut\sffamily #1};}%
}

\newcommand{\summarizeButton}[1]{%
  \tikz[baseline=(X.base)]{\node[rounded corners=2pt, fill=primary_button,
    inner xsep=3pt, inner ysep=0pt] (X) {\strut #1};}%
}

\newcommand{\applyButton}[1]{%
  \tikz[baseline=(X.base)]{\node[rounded corners=2pt, fill=secondary_button,
    inner xsep=3pt, inner ysep=0pt] (X) {\strut #1};}%
}

\newcommand{\icon}[2][1em]{
  \raisebox{-0.2em}{\includegraphics[height=#1]{figures/icons/#2}}
}

\newcommand{\smallicon}[2][0.6em]{
  \raisebox{0em}{\includegraphics[height=#1]{figures/icons/#2}}
}

\newif\ifdiff
\difffalse

\ifdiff
  \usepackage{xcolor}

  \newcommand{\added}[1]{\textcolor{blue}{#1}}
  \newcommand{\deleted}[1]{\textcolor{red}{\sout{#1}}}
\else
  \newcommand{\added}[1]{#1}
  \newcommand{\deleted}[1]{}
\fi

\begin{document}

\title{NaturalEdit: Code Modification through Direct Interaction with Adaptive Natural Language Representation}

\author{Ningzhi Tang}
\affiliation{
  \institution{University of Notre Dame}
  \city{Notre Dame}
  \state{IN}
  \country{USA}
}
\email{ntang@nd.edu}

\author{David Meininger}
\affiliation{
  \institution{University of Notre Dame}
  \city{Notre Dame}
  \state{IN}
  \country{USA}
}
\email{dmeining@nd.edu}

\author{Gelei Xu}
\affiliation{
  \institution{University of Notre Dame}
  \city{Notre Dame}
  \state{IN}
  \country{USA}
}
\email{gxu4@nd.edu}

\author{Yiyu Shi}
\affiliation{
  \institution{University of Notre Dame}
  \city{Notre Dame}
  \state{IN}
  \country{USA}
}
\email{yshi4@nd.edu}

\author{Yu Huang}
\affiliation{
  \institution{Vanderbilt University}
  \city{Nashville}
  \state{TN}
  \country{USA}
}
\email{yu.huang@vanderbilt.edu}

\author{Collin McMillan}
\affiliation{
  \institution{University of Notre Dame}
  \city{Notre Dame}
  \state{IN}
  \country{USA}
}
\email{cmc@nd.edu}

\author{Toby Jia-Jun Li}
\affiliation{
  \institution{University of Notre Dame}
  \city{Notre Dame}
  \state{IN}
  \country{USA}
}
\email{toby.j.li@nd.edu}

\renewcommand{\shortauthors}{Tang~\emph{et al.}}
\newcommand{\system}{\textsc{NaturalEdit}}

\begin{abstract}
Code modification requires developers to comprehend code, plan changes, articulate intent, and validate outcomes, making it cognitively demanding. While natural language (NL) code summaries offer a promising external representation of this process, existing approaches remain limited. Systems grounded in exploratory data analysis are restricted to narrow domains, while general-purpose systems enforce fixed NL representations and assume that developers can directly translate vague intent into precise textual edits. We present \system{}, which treats code summaries as interactive representations tightly linked to source code. Grounded in the \textit{Cognitive Dimensions of Notations}, \system{} introduces three key features: (1) adaptive, multi-faceted code summaries with a flexible \textit{Abstraction Gradient}; (2) interactive mapping mechanisms between summaries and code that ensure tight, structurally stable \textit{Closeness of Mapping}; and (3) intent-driven bidirectional synchronization that reduces \textit{Viscosity} during editing while preserving \textit{Visibility} and \textit{Consistency} through incremental diffs. A technical evaluation confirms \system{}'s viability, and a user study with 20 developers shows that it improves comprehension, intent articulation, and validation while increasing developers' \added{perceived} confidence and control.\looseness=-1
\end{abstract}

\ccsdesc[500]{Human-centered computing~Interactive systems and tools}
\ccsdesc[300]{Software and its engineering~Software notations and tools}

\keywords{Natural Language Programming, Code Modification, Code Summarization}


\maketitle

\section{Introduction}

Developers devote a substantial portion of their effort to code modification---altering existing programs to fit new tasks~\cite{shneiderman1979syntactic, planning2002economic}. This process is cognitively demanding: following Shneiderman and Mayers' model~\cite{shneiderman1979syntactic}, developers must first construct an ``internal semantics,'' a mental model of the program's logic; then form a modification plan, map it back to the program's formal syntax; and finally validate that the result reflects their intent. Crucially, this internal semantics remains implicit throughout: it is the cognitive target of comprehension, the basis of planning, and the standard against which validation is judged, yet it exists only in the developer's mind. Natural Language (NL) code summaries are especially promising as externalizations of this implicit model~\cite{myers2004natural}, which describes the functionality of the program in human-readable form. However, these summaries have traditionally served as static documentation for comprehension~\cite{leclair2020improved, stapleton2020human, wallace2025programmer}, leaving significant gaps in the later stages of expressing changes and validation.

While recent work has begun to make NL summaries interactive, existing approaches leave critical gaps. Early systems grounded in exploratory data analysis~\cite{liu2023wants, tian2023interactive, tian2024sqlucid} rely on rule-based pipelines constrained to narrow, well-defined operation spaces (e.g., SQL query rewriting), limiting their applicability to general-purpose programming, where code semantics and developer intent are far more diverse. 
One of the most relevant prior work, NL outlines~\cite{shi2024natural}, envisions NL as an interactive editing surface for code maintenance, but explicitly leaves this workflow unevaluated. Moreover, its design also has two fundamental limitations: it generates a single fixed hierarchical outline, with no mechanism to adjust granularity or format across different cognitive tasks, and it requires users to directly rewrite outline text to express changes without system mediation, sidestepping the ``fuzzy abstraction matching'' problem~\cite{sarkar2022like}, namely the difficulty of translating vague intent into the precision required to modify a representation.
More broadly, analyzing existing NL-mediated systems through the \textit{Cognitive Dimensions of Notations} framework~\cite{green1989cognitive}, grounded in empirical findings~\cite{tang2025exploring, liu2023wants}, reveals four systemic gaps: a fixed \textbf{\textit{abstraction gradient}} that cannot adapt across cognitive tasks; low \textbf{\textit{closeness of mapping}}, as implicit NL-code links burden working memory; high \textbf{\textit{viscosity}} in translating intent into summary edits; and poor \textbf{\textit{visibility}} and \textbf{\textit{consistency}} when summaries are regenerated without surfacing inspectable diffs.\looseness=-1

To address these challenges, we present \system{}, a VS Code extension that treats a program's NL description as a first-class interactive representation, introducing three core features:
\begin{enumerate}[leftmargin=2em]
    \item \textit{Adaptive Multi-Faceted Representation}: dynamically adapts NL summaries in structure and granularity, creating an adaptive \textit{abstraction gradient} tailored to the developer's current task.
    \item \textit{Interactive Cross-Representation Mapping}: provides fine-grained, structurally stable links between NL segments and code, ensuring explicit \textit{closeness of mapping} throughout the workflow.
    \item \textit{Intent-Driven Bidirectional Synchronization:} allows developers to express high-level goals as free-form instructions, which the system translates into concrete NL edits and code changes, reducing \textit{viscosity} while preserving \textit{visibility} and \textit{consistency} through incremental diffs.
\end{enumerate}

We evaluated \system{} through a two-part study. First, a technical evaluation established the soundness of the approach: a benchmark confirmed the viability of our NL-mediated workflow, with performance comparable to that of direct-instruction baselines, while an expert-driven evaluation showed that \system{} consistently generates high-quality artifacts. Second, a controlled lab study with 20 experienced developers provided strong evidence of the usability and cognitive benefits of \system{}.

In summary, we make two contributions: (1) \system{}, a VS Code extension that operationalizes NL-centric code modification \added{by integrating three interaction features into a coherent comprehension-intent-validation workflow, each mechanism addressing a usability obstacle in prior systems and} grounded in \textit{Cognitive Dimensions} analysis; and (2) a mixed-methods evaluation showing that the approach is technically viable and improves developers' \added{perceived} comprehension, intent specification, sense of control over AI-generated edits and overall modification workflow.

\section{Background and Related Work}

\subsection{Cognitive Demands of Code Modification}
\label{sec:cognitive_demands}

Modifying existing code is central to software engineering, yet it remains costly~\cite{detienne2001software, shneiderman1979syntactic}: constructing an internal semantics of existing code alone can consume up to 58\% of working time~\cite{xia2017measuring}. Developers must then bridge the \textit{Gulf of Execution}~\cite{hutchins1986direct, norman2013design} by translating intended changes into precise formal syntax, requiring substantial planning and syntax mapping effort~\cite{wirth1971program, robillard2009makes, piccioni2013empirical}. Finally, they must cross the \textit{Gulf of Evaluation} by verifying that modifications reflect their intent without introducing regressions or unintended side effects~\cite{brooks1995mythical, bacchelli2013expectations, myers2004art}. Large language models (LLMs) have transformed but not eliminated these bottlenecks. The rise of ``vibe coding''~\cite{sarkar2025vibe}, where developers delegate implementation through conversational prompts, often bypasses deep comprehension, compressing intent into a prompt and amplifying the ``fuzzy abstraction matching'' problem~\cite{sarkar2022like}. Additionally, validating AI-generated edits overwhelms cognitive load~\cite{mozannar2024reading, tang2024developer}, pushing developers toward surface-level execution checks rather than semantic understanding~\cite{tang2025exploring}. Together, these pressures disrupt the continuity across comprehension, planning, and validation, underscoring the need for NL interactions that reintegrate rather than further fragment these cognitive workflows.

\subsection{NL-Mediated Code Modification}
\label{sec:interactive_summaries}

Early \added{NL-mediated systems for} exploratory data analysis, including GAM~\cite{liu2023wants}, \textsc{Steps}~\cite{tian2023interactive}, and \textsc{SQLucid}~\cite{tian2024sqlucid}, show that translating AI-generated code back into NL can increase developer confidence and reduce ambiguity, but remain restricted to semantically constrained operation spaces. 
In general-purpose programming, Shi~\emph{et al.}~\cite{shi2024natural} introduced NL outlines, which summarize code as structured NL statements at a single fixed granularity. Their \textit{Finish Changes} prototype allows developers to drive code updates by directly rewriting outline text to express desired changes, \added{but} this editing workflow was explicitly left out of \added{their} evaluation. Beyond these, it provides only coarse navigation between NL and code, and synchronizes changes via full regeneration rather than incremental diffs that isolate what has semantically changed. \textsc{PInG}~\cite{di2025enhancing} adopts a similar premise in a more constrained setting, anchoring NL to fixed statement-level inline comments, which further limits the range of cognitive tasks it can support. Both systems treat NL as a direct editing surface: \added{to express a change, the developer must already translate a vague goal into precise summary text, confronting the fuzzy abstraction matching problem~\cite{sarkar2022like} without system support.}

Recent empirical work by Tang~\emph{et al.}~\cite{tang2025exploring} documents the concrete usability gaps in this interaction. In a study of 15 developers using \textsc{Pasta}, a minimal summary-mediated prototype, they showed that although modifying summaries can prompt code changes, the workflow exhibits specific shortcomings, including overly rigid representations, implicit NL-code links, and disproportionate editing effort. \system{} addresses these issues by \added{treating} NL as the primary artifact through which developers externalize and refine their modification intent before any code is changed, while also mapping code changes back into the NL summary \added{for validation}. The interaction design principles behind this approach are grounded in the \textit{Cognitive Dimensions of Notations} (Section~\ref{sec:design_goals}).
\section{\system{}}

\begin{figure*}[t]
    \centerline{\includegraphics[width=0.76\textwidth]{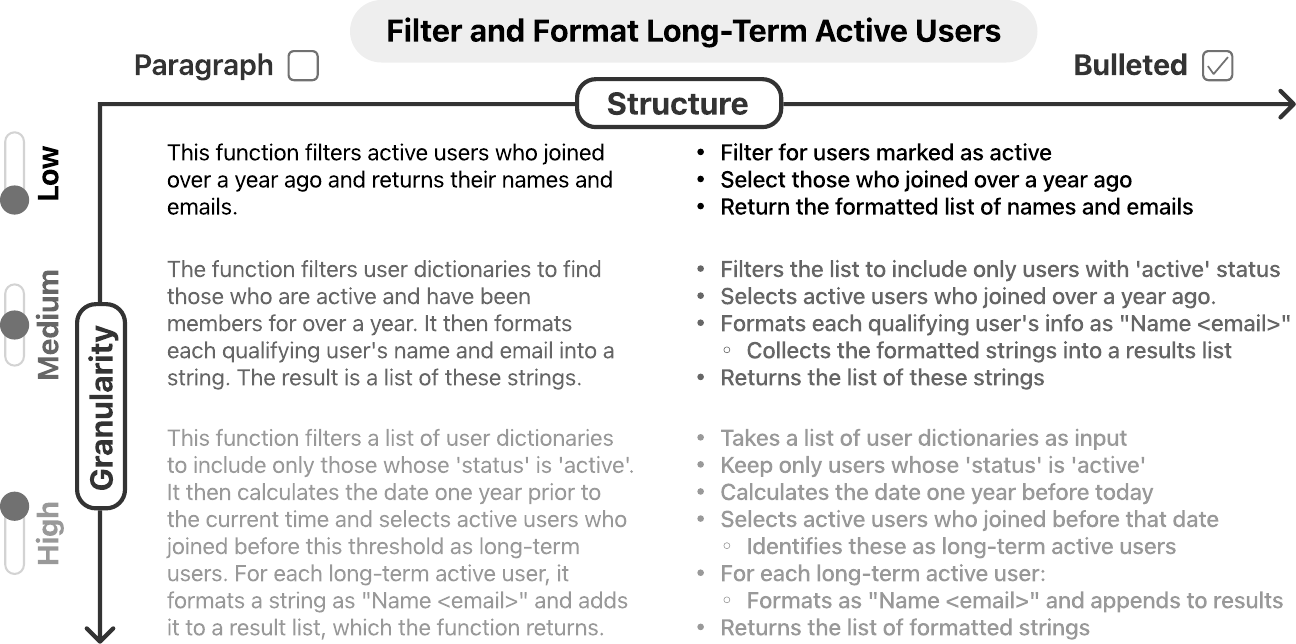}} 
    \Description{A 3 by 2 grid comparing \system{}'s natural language representations across three granularity levels (Low, Medium, High) and two structure formats (Paragraph and Bulleted), using the example function ``Filter and Format Long-Term Active Users.'' Low granularity shows a single sentence or 2 to 3 bullet points summarizing what the function does. Medium granularity adds procedural detail across 2 to 3 sentences or 3 to 5 bullets. High granularity provides a line-by-line explanation in a dense paragraph or nested bullet list. The grid is headed by UI control labels: a toggle switch for Structure and a three-stop slider for Granularity.}
    \caption{\system{} generates a concise \captionTitle{Title} and an NL representation adjustable along two dimensions:\\ \captionBox{Structure} (\icon{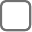}~paragraph / \icon{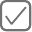}~bulleted), and \captionBox{Granularity}~(\icon{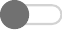}~overview, \icon{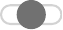}~procedural, \icon{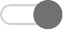}~line-by-line).}
    \label{fig:structure_granularity}
\end{figure*}

\subsection{Design Goals}
\label{sec:design_goals}

To derive our design goals, we use the \textit{Cognitive Dimensions of Notations} framework~\cite{green1989cognitive} as an analytical lens, scoped specifically to the NL-mediated code modification workflow. Rather than applying the framework exhaustively, we focus on dimensions that surface consistently as key constraints across prior studies of summary-mediated interaction~\cite{tang2025exploring, liu2023wants, shi2024natural}, as characterized in Section~\ref{sec:interactive_summaries}. \added{We first synthesize these empirically recurring failures, then use the framework's vocabulary to explain \emph{why} they} arise at the level of cognitive mechanism, and to derive principled design goals that address them. \added{Because the framework serves analysis rather than generation, the mapping between dimensions and features is not one-to-one: e.g., intent-driven synchronization addresses both viscosity and visibility.} The analysis was iteratively refined through discussions with two external experts with experience in software engineering research and developer tool design.\looseness=-1

\textbf{DG1. Enable an Adaptive Abstraction Gradient.}
Effective code modification requires developers to move fluidly across levels of abstraction~\cite{victor2011up}: from high-level orientation during comprehension, to fine-grained inspection during validation. A system enforcing a fixed \textit{abstraction gradient} cannot support this range of cognitive tasks, and prior studies confirm this as a systemic failure in NL-mediated workflows~\cite{tang2025exploring}. Prior systems enforce such rigidity in both NL representation (e.g., \textsc{PInG}~\cite{di2025enhancing} is anchored to statement-level comments; NL outlines~\cite{shi2024natural} remain static once generated) and code scope selection, which offers no awareness of the surrounding structural hierarchy. \system{} must therefore support dynamic adjustment of the NL representation, with complementary support for fluidly navigating the code scope.

\textbf{DG2. Ensure Tight and Interactive Mapping.} 
Mentally tracing relationships between NL concepts and code logic places a high load on working memory~\cite{kirsh2010thinking}, yet prior systems leave this burden entirely on the developer~\cite{tang2025exploring}: most either hide code entirely~\cite{liu2023wants} or provide only coarse, non-interactive links~\cite{shi2024natural}. Furthermore, any such links must remain valid as code evolves through edits, requiring mapping mechanisms grounded in the code's underlying structure. \system{} must therefore externalize NL-code relationships as explicit, fine-grained interactive links that remain structurally stable throughout the modification workflow.

\textbf{DG3. Minimize Viscosity in Intent Specification.}
We define \textit{viscosity} as the cognitive and physical effort required to translate high-level intent into concrete summary edits. This effort is structural, not incidental. Prior systems assume that developers can directly edit a fixed NL representation to express desired changes~\cite{shi2024natural, liu2023wants, di2025enhancing}, thereby sidestepping the fuzzy abstraction matching problem that makes intent articulation cognitively demanding. Consequently, the NL layer does not mediate between a developer's vague goal and the precise textual edits required to implement it, resulting in high viscosity for non-trivial modifications. \system{} therefore treats intent specification as a system-mediated process, allowing developers to express high-level goals naturally, while the system translates them into concrete NL edits.\looseness=-1

\textbf{DG4. Maintain High Visibility and Consistency.} 
Opaque or inconsistent updates disrupt developers' mental models and undermine trust in AI-generated changes~\cite{amershi2019guidelines, sarkar2022like, tang2025exploring}. Prior systems either discard NL after each interaction~\cite{di2025enhancing} or preserve it without exposing inspectable diffs~\cite{shi2024natural}, leaving developers without a reliable basis for validation. Full regeneration worsens this problem: when the entire summary is rewritten, developers cannot distinguish intended changes from incidental rewording. \system{} therefore presents NL updates as explicit, incremental diffs that isolate only what changed, allowing the summary to remain a persistent basis for validation throughout the modification workflow.

\subsection{Key Features}
\label{sec:key_features}

\system{} introduces three key features, organized by role: the NL representation design (Section~\ref{sec:adaptive_nl}), the mapping between representations (Section~\ref{sec:cross_mapping}), and the end-to-end modification workflow (Section~\ref{sec:overall_workflow}). 

\subsubsection{Adaptive Multi-Faceted Representation}
\label{sec:adaptive_nl}

To meet \textbf{DG1}, \system{} enables NL representation dynamically adjustable along two orthogonal dimensions: \textit{Structure} and \textit{Granularity} (Figure~\ref{fig:structure_granularity}).\looseness=-1

\textit{Structure} toggles between prose paragraphs and bulleted lists. Prior work shows that list-style presentations better support procedural understanding~\cite{karreman2007paragraphs}. Tang~\emph{et al.}~\cite{tang2025exploring} also found that developers preferred bullet lists to reason about code behavior and task decomposition, as they mirror code's logic. By contrast, paragraphs support a more holistic, narrative view. Offering both allows developers to match the format to their current cognitive task.

\textit{Granularity} controls the level of detail, ranging from a high-level overview (what a function does), to a procedural summary (how it works), to a line-by-line explanation. This progression is analogous to semantic zooming~\cite{perlin1993pad, bederson1994pad++}, reflecting developers' preference for adjustable detail to balance efficiency and thoroughness~\cite{tang2025exploring}.

Developers interact with these through simple UI controls: a toggle switch~\icon{checked.pdf} for Structure and a three-stop slider~\icon{gran_medium.pdf} for Granularity. Internally, \system{} generates a concise \captionTitle{Title} and all six representations (2 structures $\times$ 3 granularities) on demand from a single structured LLM prompt, enabling instantaneous, consistent switching.

\begin{figure*}[htbp]
    \centerline{\includegraphics[width=0.92\textwidth]{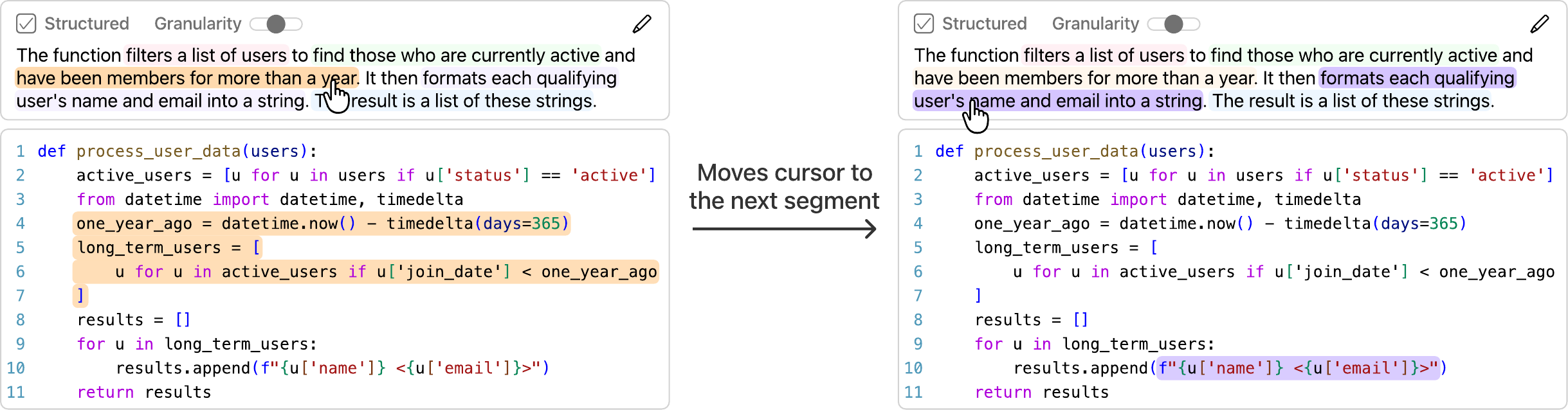}} 
    \Description{Two side-by-side screenshots of the \system{} interface illustrating interactive cross-representation mapping. Both panels show the same Python function \texttt{process\_user\_data} (11 lines) alongside its natural language summary. In the left panel, the cursor hovers over the summary segment ``have been members,'' and lines 4 to 6 of the code, which compute \texttt{one\_year\_ago} and filter \texttt{long\_term\_users}, are highlighted. In the right panel, the cursor has moved to the segment ``formats each,'' and lines 9 to 10, containing the for-loop appending formatted strings, are highlighted instead. An arrow between the panels labeled ``Moves cursor to the next segment'' illustrates the dynamic update.}
    \caption{Interactive cross-representation mapping in \system{}. As the user's \icon{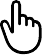}~cursor moves from one semantic segment (e.g., {\setlength{\fboxsep}{1pt}\colorbox{orange_highlight!80}{have been members...}}) to another (e.g., {\setlength{\fboxsep}{1pt}\colorbox{purple_highlight!80}{formats each...}}), the system dynamically updates the corresponding code highlight.}
    \label{fig:interactive_mapping}
\end{figure*}

\begin{figure*}[htbp]
    \centerline{\includegraphics[width=0.96\textwidth]{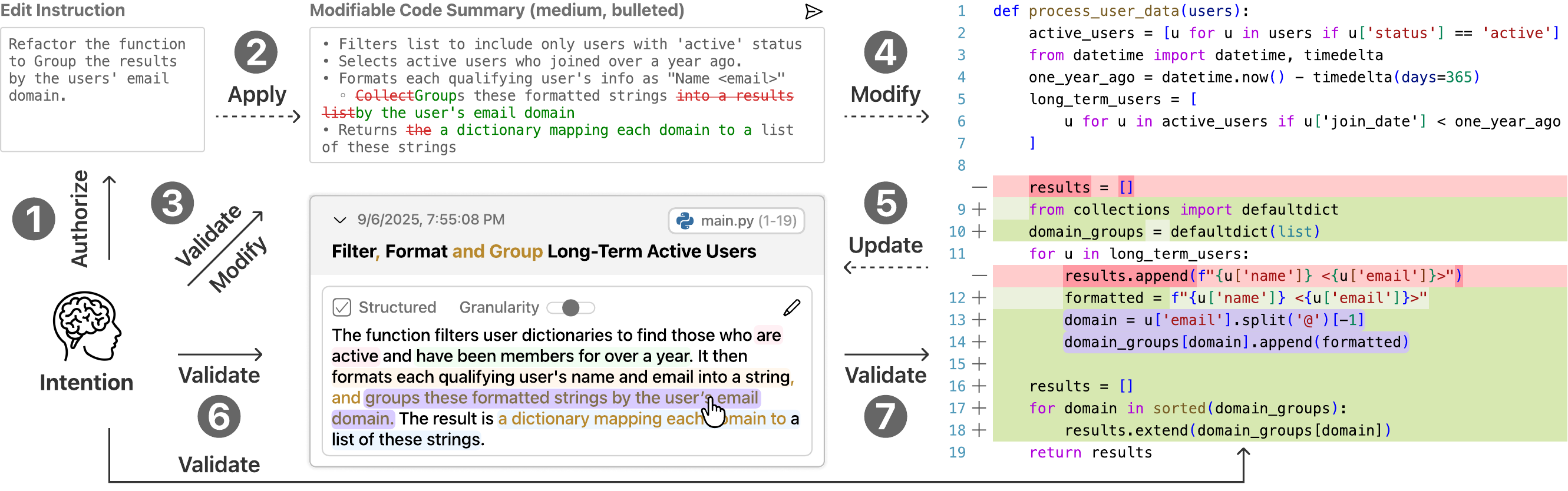}} 
    \Description{A workflow diagram of \system{}'s intent-driven bidirectional synchronization, showing seven numbered steps connected by solid arrows representing developer actions and dashed arrows representing system actions. Step 1: the developer issues an edit instruction to group results by email domain. Step 2: the system applies the instruction to the Modifiable Code Summary, generating a proposed natural language diff with insertions shown in green and deletions shown in red. Step 3: the developer validates or refines the natural language diff. Step 4: upon approval, the system modifies the code, shown as a 19-line diff with additions and deletions. Step 5: the system auto-syncs the code back to an updated natural language representation. Steps 6 and 7: the developer performs global and local validation using the side-by-side summary and code diffs, with purple highlights on changed regions.}
    \caption{Intent-driven bidirectional synchronization in \system{}: developer actions~\smallicon{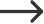} (solid arrows) and system actions~\smallicon{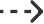} (dashed arrows) form a cohesive loop connecting intent articulation/refinement, code modification, and validation.}
    \label{fig:intent_sync}
\end{figure*}

\textit{Fluid Code Scope Navigation.} DG1 also requires fluid navigation across code scopes. Instead of requiring repeated manual selection, \system{} supports two complementary interactions: developers can move upward to enclosing scopes by clicking a code region and selecting from its AST path (e.g., method, class), or move downward by clicking a summary segment to select the corresponding code region (Section~\ref{sec:cross_mapping}). Both interactions are enabled by the Structural Alignment Engine (Section~\ref{sec:structural_alignment_engine}).

\subsubsection{Interactive Cross-Representation Mapping}
\label{sec:cross_mapping}

To address \textbf{DG2}, \system{} replaces the implicit relationship between summary and code with explicit, dynamic links. As shown in Figure~\ref{fig:interactive_mapping}, these links are fine-grained and responsive. \system{} segments each NL representation into semantic units. Hovering over a segment immediately highlights the corresponding code block; clicking navigates to and selects that scope, decoupling passive comprehension from active navigation.
Interactive mappings are generated in parallel for each summary variant. The model receives source code annotated with 1-based line numbers and returns a JSON array of (\texttt{summaryComponent}, \texttt{codeSegments}) pairs. Each \texttt{codeSegments} holds an array of objects, each specifying a code fragment and its line number in the annotated source. Line numbers are essential for precise alignment, especially when identical code fragments appear multiple times. 

Critically, these links must remain valid as the code evolves from manual edits, which simple text-based anchoring cannot guarantee. \system{} addresses this limitation through its Structural Alignment Engine (Section~\ref{sec:structural_alignment_engine}), which converts volatile line-based references into persistent AST anchors tied to structural lineage rather than physical position. As a result, tracing between summary and code becomes a low-effort perceptual operation that remains stable throughout the modification lifecycle.

\subsubsection{Intent-Driven Bidirectional Synchronization}
\label{sec:overall_workflow}

The final key feature integrates adaptive representation and interactive mapping into a cohesive end-to-end modification workflow (Figure~\ref{fig:intent_sync}).

\textit{Intent Articulation and NL Refinement.}
\icon{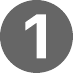}~The developer issues a high-level instruction. \icon{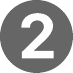}~The system proposes a summary diff (\texttt{\textcolor{insertion_diff}{insertions}} in green and \texttt{\textcolor{deletion_diff}{\sout{deletions}}} in red) as an editable intermediate artifact. \icon{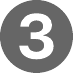}~The developer validates or refines this NL diff to ensure it precisely captures their intent before any code is modified. This directly addresses \textbf{DG3}: a single high-level command suffices for conceptually simple changes, with the intermediate diff making the system's interpretation explicit and correctable.

\begin{figure*}[htbp]
    \centerline{\includegraphics[width=0.81\textwidth]{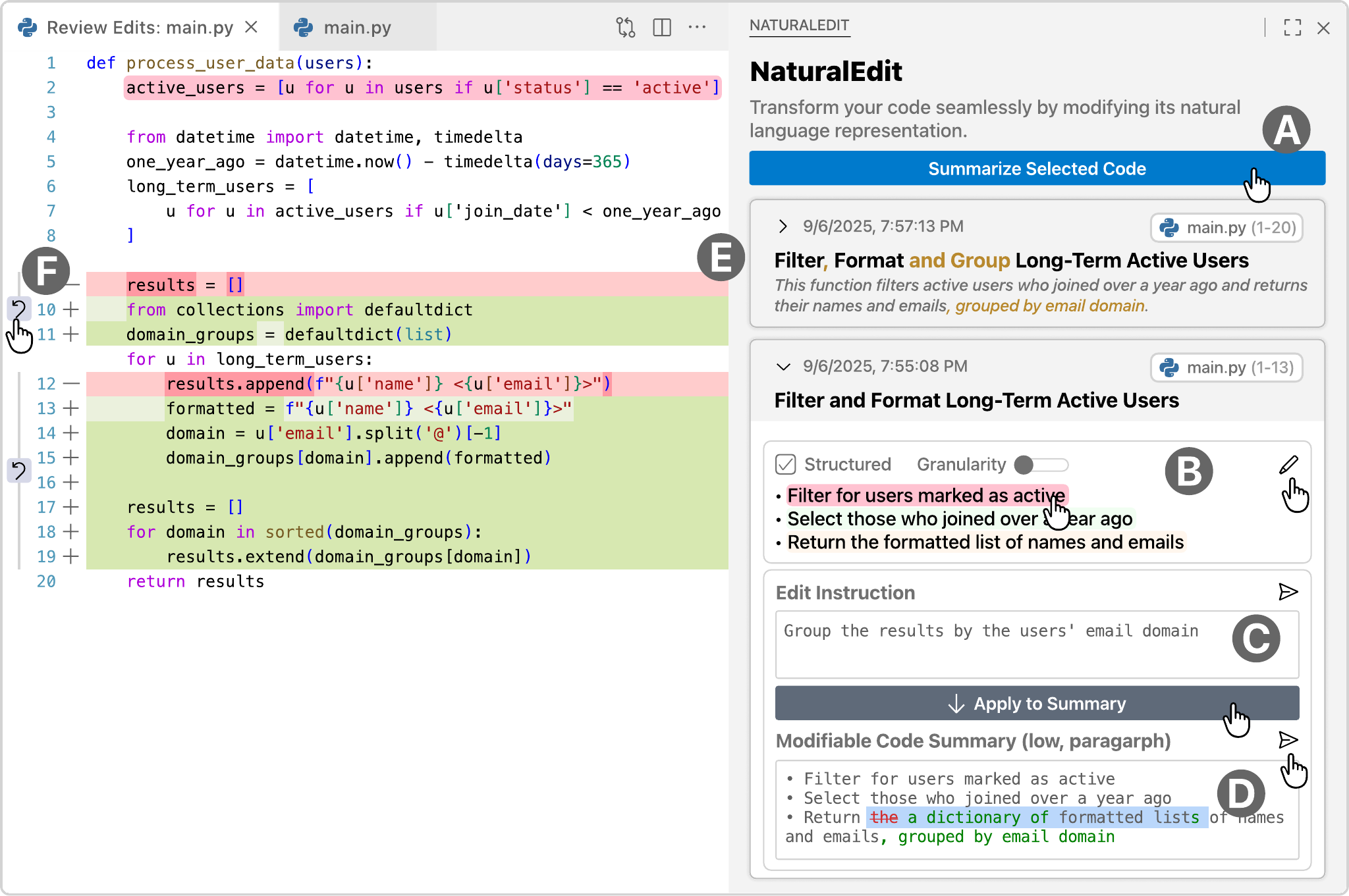}} 
    \Description{An annotated screenshot of the \system{} VS Code extension interface during the example workflow, with six labeled steps A through F. Panel A shows the ``Summarize Selected Code'' button being clicked. Panel B shows the natural language summary displayed in low-granularity bulleted format, with the developer hovering over a bullet point. Panel C shows a typed edit instruction: ``Group the results by the users' email domain.'' Panel D shows the Modifiable Code Summary with a proposed natural language diff containing strikethrough deletions and inserted text for the developer to validate. Panel E shows the synchronized code diff in the editor, with added and removed lines highlighted in a Review Edits view. Panel F shows a revert cursor hovering over an unintended change in the code diff.}
    \caption{The \system{} interface during the example workflow, showing key steps: summarizing code~(A), interacting with the NL representation~(B), issuing an instruction~(C), validating the NL diff~(D), reviewing the synchronized code diff with Incremental Diffs~(E), and reverting unintended changes~(F).}
    \label{fig:example_workflow}
\end{figure*}

\textit{Code Modification and Bidirectional Synchronization.}
\icon{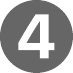}~Upon approval, the system generates the code change by providing the LLM with the original code alongside both summary versions, directing attention to the developer-approved NL diff. \icon{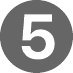}~\system{} then automatically regenerates the NL summary as \textcolor{incremental_diff}{\textbf{\textsf{Incremental Diffs}}} (\added{targeted} edits that isolate the semantic impact), keeping the summary synchronized with code. \added{When the developer instead edits the code directly, the code is treated as the source of truth, and background validity checking (Section~\ref{sec:structural_alignment_engine}) re-resolves the mapping accordingly.} 

\textit{Multi-Level Validation.}
\icon{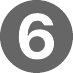}~Developers perform \textit{global validation} by inspecting the summary and code diffs side by side. \icon{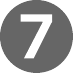}~\textit{Local validation} then leverages interactive mapping directly on highlighted diff regions (e.g., {\setlength{\fboxsep}{1pt}\colorbox{purple_highlight!80}{purple highlights}} in Figure~\ref{fig:intent_sync}), enabling fine-grained verification of NL-code correspondence. This dual-diff loop addresses \textbf{DG4} by ensuring visibility and consistency, giving developers confidence that the final program reflects their intent.\looseness=-1

\subsection{Example Workflow}

To illustrate how \system{} integrates its features, we follow Grace as she performs the refactoring task in Figure~\ref{fig:example_workflow}. Her goal is to modify the function \texttt{\textcolor{function_keyword}{process\_user\_data}}, which \textit{\textsf{filters active users who joined over a year ago and returns their names and emails}}, so that it groups the resulting users by email domain. \icon{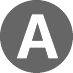}~Grace begins by selecting \texttt{\textcolor{function_keyword}{process\_user\_data}} in the editor and clicking \summarizeButton{\textcolor{white}{\textsf{\small Summarize Selected Code}}}, which opens a new interactive session.

\icon{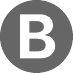}~Grace then switches to a structured, low-granularity view, which presents the function's logic as three bullet points for quick comprehension. She hovers over each point to highlight the corresponding code, building a mental model before deciding an edit scope. She then clicks the edit button~\icon{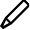} to load the summary into the \textsf{\textbf{Modifiable Code Summary}} area.
\icon{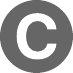}~Rather than editing the summary directly, Grace types \texttt{Group the results by the users' email domain} into the \textsf{\textbf{Edit Instruction}} box and clicks \applyButton{\textcolor{white}{\textsf{\small $\downarrow$ Apply to Summary}}}. The system generates a proposed change, inserting the phrase \texttt{\textcolor{insertion_diff}{grouped by email domain}} into the summary. 
\icon{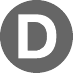}~Reviewing the proposed diff, she refines it to additionally specify the return type as {\setlength{\fboxsep}{1pt}\colorbox{mouse_selection}{\texttt{\textcolor{deletion_diff}{\sout{the}} \textcolor{insertion_diff}{a dictionary of} formatted list\textcolor{insertion_diff}{s}}}}, then approves by clicking~\icon{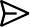}.
\icon{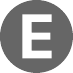}~\system{} applies the change and presents a synchronized code diff, while generating a new session with updated NL representations and \textcolor{incremental_diff}{\textbf{\textsf{Incremental Diffs}}}. Grace confirms that \texttt{collections.defaultdict} is correctly imported and that the grouping logic matches her intent. \icon{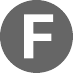}~If the system introduces incorrect logic, she can revert~\icon{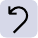} specific lines in the diff view or continue iterating in a new summary session. \added{All sessions remain live while the IDE is open, so Grace can later revisit \texttt{\textcolor{function_keyword}{process\_user\_data}} without re-summarizing.}

\subsection{Implementation}
\label{sec:system_implementation}

\system{} is implemented as a VS Code extension\footnote{\url{https://code.visualstudio.com/api}}, with its main UI rendered in a side panel. This architecture ensures compatibility with all forks of VS Code, e.g., Cursor. The front end is built with React and Vite inside a VS Code Webview and styled with \texttt{@vscode/webview-ui-toolkit} for seamless integration into different themes. All NL processing is powered by OpenAI's GPT-4.1\footnote{\url{https://openai.com/index/gpt-4-1/}}; prompt templates are provided in Appendix~\ref{sec:prompt_templates}. NL summary diffs are computed client-side using \texttt{diff-match-patch} library\footnote{\url{https://github.com/google/diff-match-patch}}; code diffs use VS Code's native \texttt{vscode.diff} engine, and editor highlights are rendered with \texttt{TextEditor.setDecorations}. 

\subsubsection{AST-Driven Structural Alignment Engine}
\label{sec:structural_alignment_engine}

To maintain alignment robustness at both the session and mapping levels as code evolves, \system{} replaces volatile line-number and text-based references with persistent \textbf{\textit{AST Anchors}}: structural descriptors that identify where each summary segment maps to in the code, parsed via Tree-sitter\footnote{\url{https://tree-sitter.github.io/tree-sitter/}}. Each anchor encodes a location through two components: (1) a \textit{structural path} from the file root to the target node, recording the AST type, name, and sibling index at each level; and (2) \textit{target identity}, capturing the AST type, name, and content hash of the minimal AST node enclosing the target code scope (e.g., a function named \texttt{process\_user\_data}). When code is edited, the system re-resolves each anchor by scoring candidates as $s_{\text{path}} \cdot s_{\text{target}}$, traversing the path top-down and matching each level by type, then name, then position, so refactored nodes can still be located \added{and moves among siblings resolve correctly, as position serves only as the final tie-breaker}. This keeps interactions focused on code semantics: minor formatting changes, such as added whitespace or blank lines, do not affect the structural path and therefore leave all anchors intact, while meaningful edits that move or rename a code scope are gracefully re-resolved rather than silently misaligned.
A summary session remains \textit{Valid} as long as its anchors can be successfully re-resolved; otherwise, it is flagged as \textit{Outdated} in the UI, prompting the developer to refresh.
This mechanism supports all Tree-sitter languages, including Python and JavaScript; unsupported languages fall back to fuzzy text matching.
\section{Technical Evaluation}

We conducted a two-phase technical evaluation: a benchmark assessment verifying that NL-mediated modification preserves accuracy comparable to direct-instruction baselines, and an expert evaluation of the intermediate NL artifacts (summaries, mappings, and diffs) central to the user experience.

\subsection{Code Editing Benchmark Evaluation}
\label{sec:technical_evaluation_benchmark}

\subsubsection{Experimental Setup}

We evaluated on two standard Python code editing benchmarks, both providing NL instructions \added{and unit tests for correctness validation.} \textit{CanItEdit}~\cite{cassano2024can} \added{provides 105 hand-crafted tasks, each in two instruction styles: \textit{Lazy} instructions are brief and potentially ambiguous, resembling human-authored requests, while \textit{Descriptive} instructions specify the intended change in detail.} \textit{EditEval}~\cite{li2024instructcoder} provides 194 tasks from GitHub commits, HumanEval~\cite{chen2021evaluating}, and MBPP~\cite{austin2021program}. We compared \textit{Direct Instruction} (single-step code generation from instruction) against \textit{NL Mediation} (\system{}'s two-step workflow across six representation variants). Both conditions used GPT-4.1 with prompts based on \system{} (Appendix~\ref{sec:prompt_templates}), with minor adjustments for the benchmark format (e.g., removing unavailable file context). We report \textit{Pass@1}, the proportion of tasks where the first generated code passes all \added{of the benchmark's} ground-truth unit tests.

\subsubsection{Results}

Figure~\ref{fig:datasets_result} (Appendix~\ref{sec:benchmark_evaluation_results_appendix}) \added{reports Pass@1 for all six representation variants on each benchmark; here we summarize the aggregate patterns.}
\textbf{The NL-mediated workflow preserves technical soundness.} \system{} achieves performance broadly comparable to the Direct Instruction baseline across all benchmarks (average $-1.92\%$, maximum $-8.98\%$), suggesting that an intermediate NL layer does not fundamentally compromise code correctness.
\added{We interpret the modest decrease as the cost of the} intermediate NL representation, which requires the model to summarize and reinterpret both the code and the instruction. In most cases, errors arose when this representation omitted or simplified key semantics, such as boundary conditions or constraints needed for correct edits. This suggests a trade-off between NL readability and the semantic precision required for accurate code modification.

\textbf{NL mediation is more robust to vague instructions.} Compared with the baseline, NL mediation performs better on \textit{Lazy} instructions (+0.48\%) than on \textit{Descriptive} ones (-4.45\%). This suggests that grounding vague intent in an explicit NL representation reduces the risk of underspecification.

\subsection{Expert Ratings of Intermediate Artifacts}
\label{sec:technical_evaluation_quality}

\subsubsection{Evaluation Protocol}

We used the six code modification tasks from our user study (Section~\ref{sec:tasks}) as the corpus, as they involve multi-file edits and better reflect the real-world scenarios \system{} is designed to support. For each task, we generated six representation variants for both the initial code and the ground-truth final solution; final-solution summaries were produced as incremental diffs against the original. This yielded a corpus of 72 summaries (36 with diffs) and 487 summary-code mappings.

Drawing on prior work in code summarization~\cite{treude2020beyond, wallace2025programmer}, the two authors collaboratively defined criteria for each artifact: \textit{Accuracy} (correctness of description) and \textit{Clarity} (readability) for summaries; \textit{Segmentation Granularity} (appropriateness of segment size), \textit{Accuracy} (correctness of NL-code links), and \textit{Coverage} (completeness of links) for mappings; and \textit{Faithfulness} (accuracy of diff representation), \textit{Completeness} (capture of all semantic changes), and \textit{Salience} (prominence of important changes) for diffs. The full rating guide is provided in Appendix~\ref{sec:expert_rating_guide}.

Two computer science researchers, one specializing in AI model evaluation and the other in developer tools, conducted the assessment. They independently scored all items on a 5-point scale (1 = Very Poor, 5 = Excellent).
We assessed inter-rater reliability using percent agreement, which is better suited than Cohen's kappa for highly skewed data~\cite{feinstein1990high}, e.g., when the majority of ratings are 5. We report both strict (exact-match) and relaxed (within $\pm 1$ point) agreement: 70.4\% (96.3\%) for Summary Quality, 74.1\% (99.1\%) for Diff Quality, and 94.3\% (98.0\%) for Mapping Quality. Final scores were determined by averaging ratings that differed by one point and resolving larger disagreements through consensus. These discussions also informed a qualitative analysis of recurring error patterns.\looseness=-1

\subsubsection{Results}
\label{sec:expert_rating_results}

Figure~\ref{fig:ratings_result} (Appendix~\ref{sec:expert_rating_results_appendix}) shows \added{ratings for all eight criteria across the six variants}.
\textbf{The generated NL artifacts were consistently rated as high quality.} The majority (42/48) of mean scores fell between 4.5 and 5.0 on a 5-point scale, confirming the viability of our pipeline for producing usable NL representations.\looseness=-1

\textbf{However, segmentation utility decreased at high granularity in unstructured summaries.} Raters scored \textit{Segmentation Granularity} significantly lower for the high-granularity unstructured variant ($M=4.08$, $SD=0.74$; $W=2.5$, $p=.038$). Raters observed that, at this level of detail, consecutive highlighted blocks often corresponded to nearly identical code regions, suggesting that coarser semantic segmentation would be more effective.

\textbf{A critical trade-off emerged between diff detail and salience.} 
While \textit{Faithfulness} and \textit{Completeness} remained highly rated, \textit{Salience} declined with increasing granularity and was lowest in the high-granularity unstructured variant ($M=3.75$, $SD=1.08$). Raters attributed this to two factors: (1) the large volume of insertions, deletions, and rewordings made important changes harder to distinguish from minor ones, and (2) action-oriented phrasing (e.g., ``\textit{Current price... added to the response}'') occasionally broke the summary's ``objective description'' metaphor.

\section{User Study}
\label{sec:controlled_user_study}

We conducted a controlled lab study to examine \system{}'s usability and cognitive impact on code modification, focusing on how its three core features shape developers' comprehension, intent specification, and validation workflows\footnote{Approved by the Institutional Review Board (IRB) at our institution.}.

\subsection{Study Design}

\subsubsection{Participants}

We recruited 20 participants (12 male, 8 female; ages 21--52, $M=27.15$, $SD=8.25$) through purposive sampling~\cite{etikan2016comparison}. All had substantial experience with Python and JavaScript. The sample included two undergraduates, 11 graduate students, and seven industry professionals, with an average of 9.55 years of programming experience ($SD=9.16$). All 13 student participants majored in Computer Science or Engineering, and eight reported prior internship experience. Each participant received a \$66 Amazon gift card.
Participants reported using LLM-powered programming tools either ``frequently'' (7) or ``almost always'' (13), including ChatGPT (18), GitHub Copilot (15), Cursor (11), and Claude Code (9). Most also had relevant domain experience in the study scenarios: web development (20/20) and machine learning (18/20). A detailed demographic summary is provided in Table~\ref{tab:demographics} (Appendix~\ref{sec:participants_appendix}).

\subsubsection{Baseline}

\added{To examine whether} \system{}'s key features \added{improve on existing NL-mediated systems}, we built a controlled baseline in which the three core interaction designs were disabled: \textit{Adaptive Multi-Faceted Representation} was replaced with a fixed, medium-level paragraph summary; \textit{Interactive Cross-Representation Mapping} was removed; and \textit{Intent-Driven Synchronization} was reduced to manual summary editing. This ablated design is functionally comparable to prior systems such as \textsc{Pasta}~\cite{tang2025exploring} and GAM~\cite{liu2023wants}, and it was implemented using the same VS Code framework and UI style as \system{} (screenshot in Figure~\ref{fig:baseline_interface}, Appendix~\ref{sec:appendix_baseline}).\looseness=-1

\subsubsection{Programming Tasks}
\label{sec:tasks}


We designed two programming tasks in full-stack web development and machine learning, implemented in JavaScript and Python, respectively. Tasks are summarized here and described in full in Appendix~\ref{sec:appendix_tasks}:

\begin{itemize}[leftmargin=1em]
    \item \textbf{Task 1.} \textit{Finance Dashboard}: A Node.js/React stock price visualization application. The subtasks included (a) implementing a helper function to format the tick labels on the x-axis, (b) adding the current price to the backend API, and (c) displaying it as a reference line on the chart.
    \item \textbf{Task 2.} \textit{MVP Predictor}: A Python pipeline for scraping NBA player data and predicting MVP rankings. The subtasks included (a) extending the scraper with advanced statistics, (b) adjusting the \texttt{n\_estimators} hyperparameter of the \texttt{XGBRanker} model, and (c) improving the type and color scheme of the visualization chart.
\end{itemize}
Task realism was validated through participant ratings, with all 20 participants rating the tasks at least 4 on a 7-point Likert scale. Across tasks, participants worked with code at varying scopes, ranging from sub-function regions to multiple related code segments across files.
To minimize bias in how participants instructed the system, we designed the task descriptions following the Natural Programming Elicitation method~\cite{myers2016programmers}. Each task was presented with annotated target visuals and brief textual descriptions, preventing the direct reuse of the task descriptions as instructions.



\subsubsection{Study Protocol}

We conducted a within-subjects user study comparing \system{} against \textit{Baseline}. Both conditions were deployed in CodeSandbox\footnote{\url{https://codesandbox.io/}}, a web-based fork of VS Code where researchers can preconfigure Node.js and Python environments, eliminating local setup issues and screen-sharing latency. The IDE theme was set to Default Light+ for visual consistency. Participants accessed the environment through provided links, with task instructions shown on a separate device to minimize context switching.

Each session lasted approximately 90 minutes. Participants first signed a consent form and completed a pre-study questionnaire on demographics and programming experience. To reduce performance anxiety and encourage authentic behavior~\cite{rubin2008handbook}, we told participants that the study evaluated the systems rather than their performance. Task and condition order were counterbalanced to mitigate learning and order effects~\cite{nielsen1994usability}. Before each condition, participants completed a guided onboarding example, and then worked on the task for 25 minutes.

We used a mixed-methods approach for data collection. The environment automatically logged all interactions to Firebase, including timestamps, feature usage events, and LLM response contexts; sessions were also screen- and audio-recorded. After each condition, participants completed the NASA-TLX~\cite{hart1986nasa} cognitive workload measure and self-reported their understanding of both the original code and the system-generated modifications.

After both tasks, participants completed a final questionnaire with parallel sections for each condition, including the UMUX-Lite usability scale~\cite{lewis2013umux} and custom items on key modification stages and \system{}'s interactive features. The session concluded with a 20-minute semi-structured interview \added{covering participants' typical modification workflows, their experiences with each of \system{}'s core features, and their perceived control over and trust in the system's output, along with follow-up questions grounded in observed behaviors.}. Full questionnaire items and the interview protocol are provided in Appendices~\ref{sec:questionnaires} and~\ref{sec:interview}.

\begin{figure*}[htbp]
    \centerline{\includegraphics[width=0.92\textwidth]{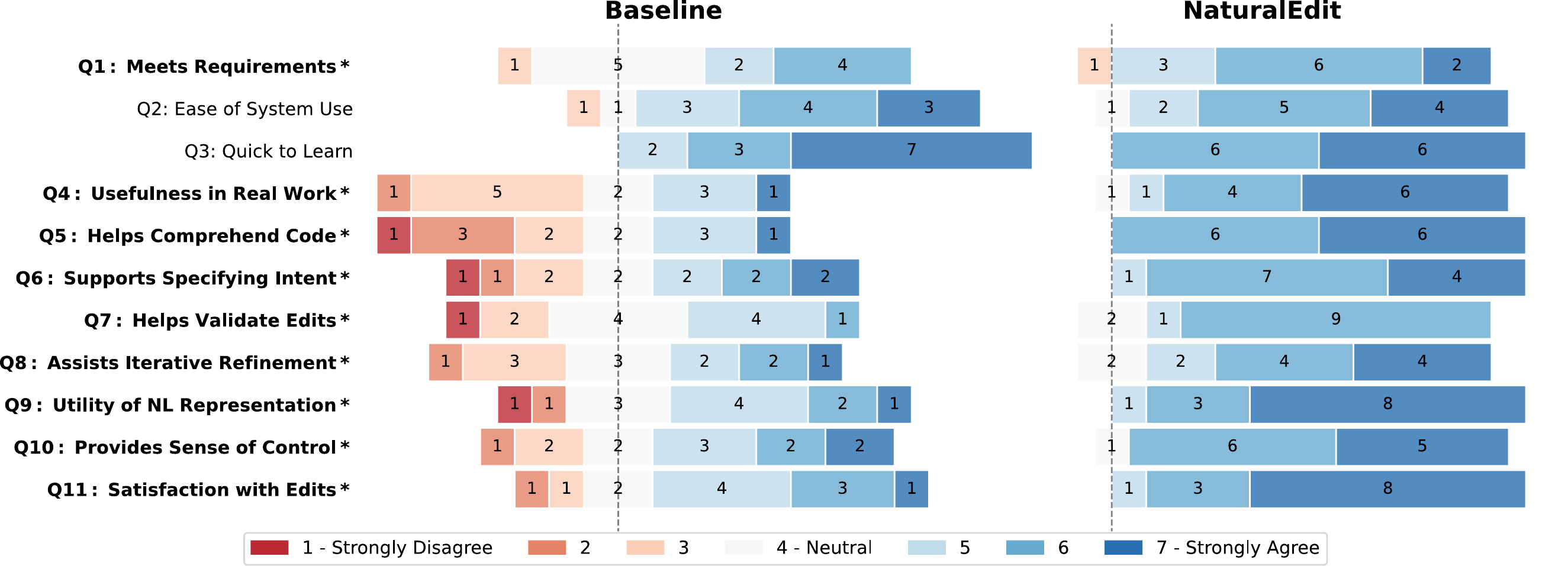}}
    \Description{A diverging stacked bar chart comparing participant ratings on 11 questions (Q1 through Q11) between Baseline and \system{} using a 7-point Likert scale, where 1 means Strongly Disagree, and 7 means Strongly Agree. Questions cover UMUX-LITE usability items (Q1 and Q2) and custom utility items (Q3 through Q11). Questions with statistically significant differences at $p < 0.01$ are marked with an asterisk. \system{} shows notably higher agreement on Q1 (Meets Requirements), Q4 (Usefulness in Real Work), Q6 (Supports Specifying Intent), Q8 (Assists Iterative Refinement), Q10 (Provides Sense of Control), and Q11 (Satisfaction with Edits).}
    \caption{Comparison of usability (Q1–Q2, UMUX-LITE) and utility (Q3–Q11) between \textit{Baseline} and \system{}. Questions with statistically significant differences ($p < 0.01$) are marked with an asterisk (*), based on Wilcoxon signed-rank tests.}
    \label{fig:likert_comparison}
\end{figure*}

\subsubsection{Data Analysis}

For qualitative analysis, we transcribed the interviews and followed an iterative, discussion-based thematic analysis process~\cite{lazar2017research, williams2019art}. One author coded 50\% of the data to develop a preliminary codebook, and a second author then coded the same subset, refining the codes through discussion. Using the unified codebook, the two authors collaboratively coded the remaining data and continuously discussed interpretations to maintain agreement. Because the analysis was discussion-based, we did not calculate a formal inter-coder reliability metric~\cite{mcdonald2019reliability}. The complete codebook is provided in Appendix~\ref{sec:codebook}.

For quantitative analysis, we used nonparametric tests appropriate for small samples. Within-subject comparisons between \system{} and \textit{Baseline} were conducted using the Wilcoxon signed-rank test ($\alpha = .05$). We report the median for each condition ($Mdn_N$ for \system{} and $Mdn_B$ for \textit{Baseline}), the test statistic $W$, and the $p$-value.
To analyze interaction behaviors, we constructed transition graphs from user logs by preprocessing raw events into higher-level actions. Two filters were applied: (1) mapping hover events with dwell times under 500 ms were excluded, and consecutive hovers were merged into \texttt{Inspect Mapping} actions; (2) consecutive \texttt{Adapt Summary Level} events were collapsed into a single action reflecting the final state. Results are shown in Figure~\ref{fig:interaction_transition}.

\subsection{Study Results}
\label{sec:user_study_results}

\begin{figure}[htbp]
    \centerline{\includegraphics[width=\columnwidth]{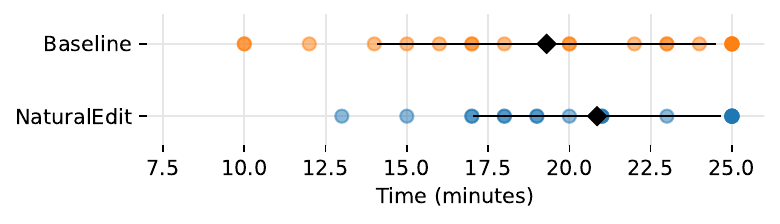}}
    \Description{A dot plot showing the distribution of time in minutes spent on tasks per participant, with one row for \system{} and one row for Baseline. The horizontal axis ranges from 7.5 to 25 minutes. Each dot represents one participant's task completion time. \system{} dots are more tightly clustered between approximately 15 and 25 minutes, while Baseline dots are more spread out, ranging from approximately 8 to 25 minutes.}
    \caption{Distribution of time spent on tasks per participant.}
    \label{fig:time_record}
\end{figure}

\system{} achieved higher task correctness than \textit{Baseline} (95.0\% vs.\ 81.7\%), while completion times were comparable (\textit{Baseline}: $19.30 \pm 5.21$ min; \system{}: $20.85 \pm 3.80$ min; $W=59.5$, $p=.256$). Despite similar time investment, behavioral data reveal systematic differences in interaction strategies (Figure~\ref{fig:interaction_transition}). Participants also self-reported significantly higher satisfaction with \system{}'s code modifications (Q11: $Mdn_N=7.0$, $Mdn_B=5.0$; $W=0.0$, $p=.003$).

\subsubsection{Overall Usability \& Experience (RQ1)}
\label{sec:rq1}

\textbf{\system{} demonstrated significantly higher usability, and \added{participants reported} they would prefer it for real-world development.} The overall SUS score for \system{} ($Mdn_N=77.07$) was significantly higher than \textit{Baseline} ($Mdn_B=66.23$, $W=34.5$, $p=.026$). Participants reported that \system{} better met their requirements ($Mdn_N=6.0$, $Mdn_B=4.5$, $W=11.0$, $p=.004$) and was significantly more useful in real development work ($Mdn_N=6.0$, $Mdn_B=4.0$, $W=8.0$, $p=.001$). Perceived ease of use and learnability were comparable across conditions ($p>.3$), suggesting that \system{} preserved the simplicity of the bare-bones \textit{Baseline} despite its additional features. Cognitive workload (NASA-TLX) showed no significant differences across all six dimensions ($p>.1$).

\begin{figure*}[htbp]
    \centerline{\includegraphics[width=0.9\textwidth]{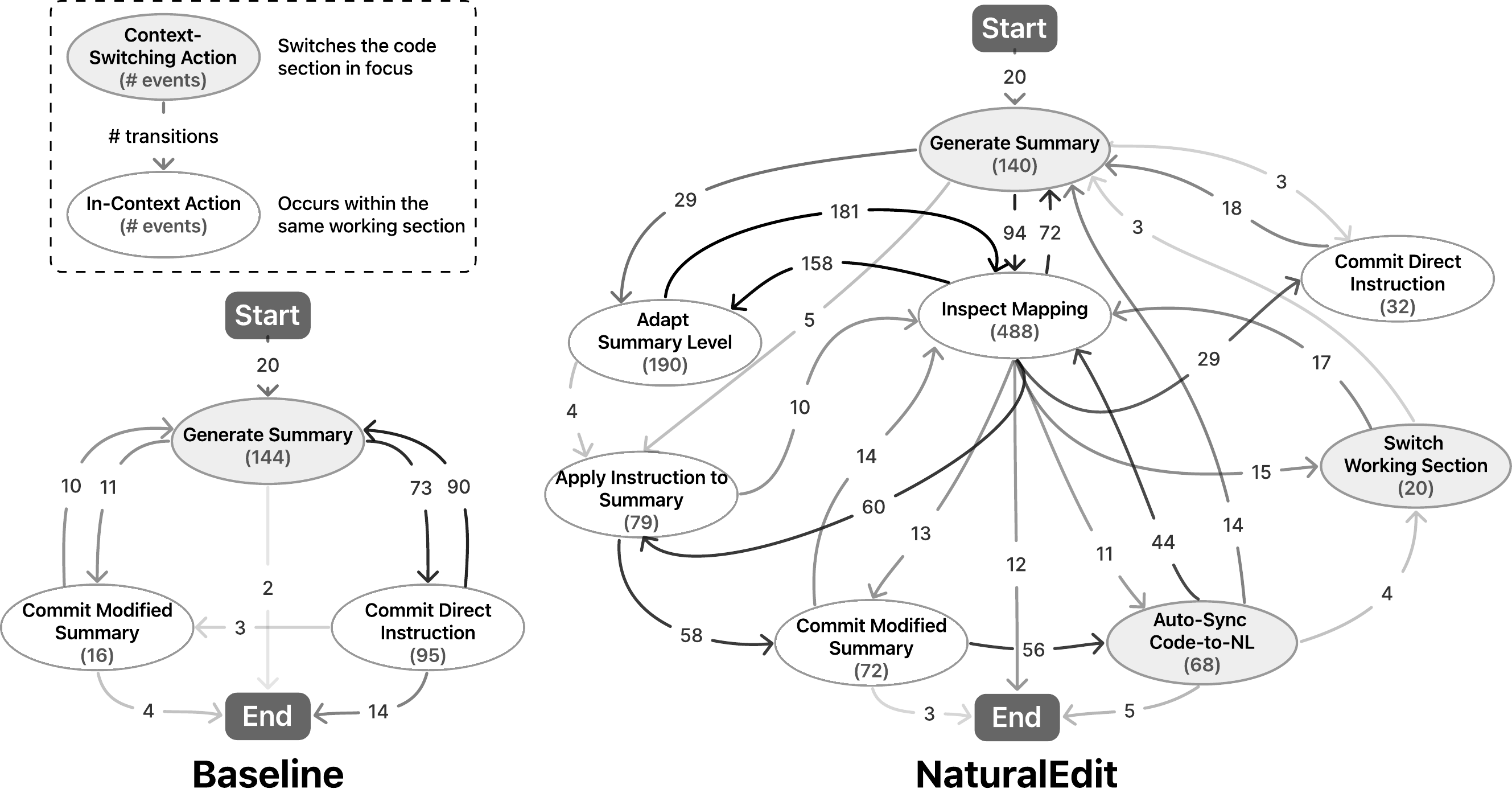}}
    \Description{Two transition graphs side by side depicting user workflows for Baseline on the left and \system{} on the right. Both graphs begin with a Start node representing 20 participants and end at an End node. Nodes represent user actions with occurrence counts in parentheses. Edges represent transitions with opacity log-scaled by frequency, where darker edges represent more common pathways. In the Baseline graph, the primary pathway leads from Generate Summary (144 occurrences) directly to Commit Direct Instruction (95 occurrences) with few detours. In the \system{} graph, additional nodes include Inspect Mapping (488), Adapt Summary Level (190), Apply Instruction to Summary (79), Commit Modified Summary (72), Auto-Sync Code-to-NL (68), Switch Working Section (20), and Commit Direct Instruction (32), with many more interconnecting transitions showing a richer, natural-language-centric workflow.}
    \caption{Transition graphs of user workflows in the \textit{Baseline} (left) and \system{} (right) conditions. Both graphs represent a complete code modification episode, with explicit start and end states. Nodes represent user actions, with occurrence counts shown in parentheses. Edges represent transitions between actions, with opacity log-scaled by transition frequency (darker represents more common pathways). For clarity, transitions that occurred only once between intermediate actions are omitted.}
    \label{fig:interaction_transition}
\end{figure*}

\textbf{\system{} shifted modification strategies toward an NL-centric workflow.} As shown in Figure~\ref{fig:interaction_transition}, 72/104 modifications were completed via the modified summary. In contrast, participants in \textit{Baseline} avoided manual summary edits entirely (95/111 modifications via direct-instruction fallback). This shift reflects reduced viscosity: participants described manual summary editing in \textit{Baseline} as ``\textit{cumbersome and taking significant effort}'' (P5).\looseness=-1

\textbf{\system{} enhanced developers' sense of control via enforced comprehension and shared grounding.} (Q10: $Mdn_N=6.0$, $Mdn_B=5.0$, $W=7.50$, $p=.008$). \system{} encouraged a ``comprehend-then-act'' strategy~\cite{shneiderman1979syntactic}, contrasting with vibe coding practices~\cite{geng2025exploring}. Participants also reported significantly better \added{perceived} understanding of AI-generated edits with \system{} ($Mdn_N=5.5$, $Mdn_B=4.5$, $W=15.0$, $p=.009$). The NL summary served as a shared ground for validating intent and system interpretation, making the ``\textit{black box more transparent}'' (P11) and giving developers ``\textit{greater confidence it would work}'' (P12).

\textbf{NL mediation diverged between declarative and procedural tasks.} Participants rated the NL representation in \system{} as significantly more useful ($Mdn_N=7.0$, $Mdn_B=5.0$, $W=0.0$, $p=.001$), with over 800 NL-mediated actions recorded. For declarative tasks (e.g., UI changes), participants preferred direct instructions, as visual feedback reduced the need for deep comprehension (P2: ``\textit{I care more about the final effect}''). For procedural tasks (e.g., backend algorithms), participants consistently used the summary-mediated path, where correctness required semantic validation beyond execution results (P12: ``\textit{strongly need additional representation to help understand changes}'').


\subsubsection{Abstraction Gradient (RQ2)}
\label{sec:rq2}

\begin{figure}[htbp]
    \centerline{\includegraphics[width=\columnwidth]{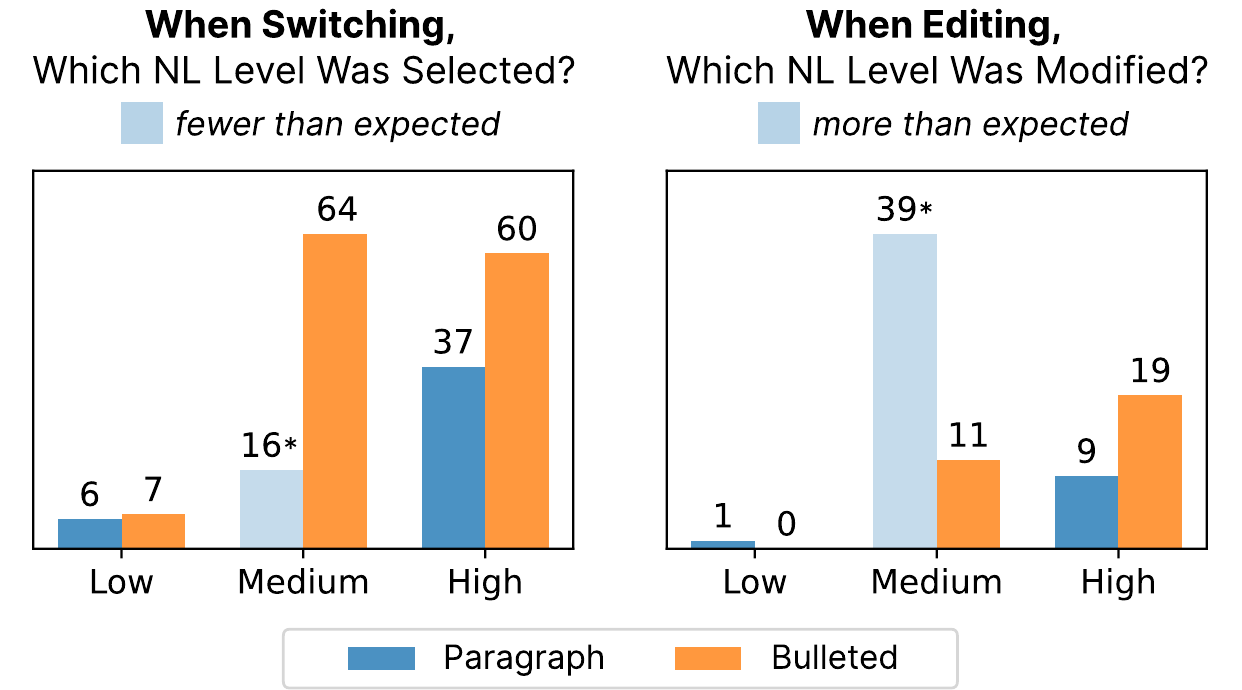}}
    \Description{Two side-by-side bar charts showing usage counts of natural language granularity levels (Low, Medium, High) and structure types (Paragraph, Bulleted) for explicit switches on the left and explicit edits on the right. In the switching chart, Medium Paragraph bars are marked with an asterisk and labeled as fewer than expected, while High Bulleted bars are the tallest. In the editing chart, Medium Paragraph bars are again marked with an asterisk and labeled as more than expected, indicating a default-view bias. High Bulleted and High Paragraph bars dominate in the editing chart.}
    \caption{Usage counts across NL granularity and structure, considering only explicit switches or edits. \textit{Medium-Paragraph} (lighter blue, *) reflects a default-view bias: underrepresented in switches but overrepresented in edits.}
    \label{fig:interaction_bar_charts}
\end{figure}

\textbf{Adaptive representation supported top-down comprehension through on-demand granularity adjustment.} \system{} significantly improved self-reported code comprehension (Q5: $Mdn_N=6.5$, $Mdn_B=4.5$, $W=0.0$, $p=.0003$), with 19 participants rating the feature 6 or 7 (Figure~\ref{fig:likert_non_comparison}). Participants appreciated starting from a ``\textit{coarser view for initial understanding}'' (P3) before drilling into details (P1, P2, P3, P8, P15, P20). However, interaction logs show they rarely switched to the lowest granularity (Figure~\ref{fig:interaction_bar_charts}), suggesting the default medium level was usually sufficient as an entry point.

\textbf{Developers generally preferred high-granularity summaries for modification, while using granularity opportunistically.} As shown in Figure~\ref{fig:interaction_bar_charts}, participants often switched to and committed edits at the high-granularity level, valuing its comprehensiveness as a basis for editing (P3, P7, P9, P10, P12, P16, P19), and its support for incremental changes that felt less likely to ``\textit{affect other parts}'' (P8, P10). Yet this preference was not rigid: for high-level changes, a concise summary likely provided sufficient scaffolding.

\textbf{Most participants preferred bulleted summaries for their procedural alignment and scannability, though paragraphs were valued for narrative coherence.} Across all granularity levels, participants more often switched to structured NL (Figure~\ref{fig:interaction_bar_charts}, left), finding it better aligned with code structure and easier to skim (P1, P3, P5, P6, P8, P10, P11, P12, P16, P17, P20). Some, however, preferred paragraphs for their logical continuity (P2, P4, P7, P8, P15), noting that bullet lists could feel ``\textit{fragmented in overall logic}'' (P7) and obscure causal relationships between sentences.

\subsubsection{Closeness of Mapping (RQ3)}

\textbf{Interactive mapping was the central action in the workflow, serving as the primary bridge between NL and code.} With 488 occurrences, it was the most frequent action in \system{} (Figure~\ref{fig:interaction_transition}), and 15/20 participants strongly agreed that it made the code-summary relationship explicit (Figure~\ref{fig:likert_non_comparison}). Mapping also played a pivotal transitional role: it was the most common first action after generating a summary (94/140) or switching sections (17/20), and the most common action both before (158/190) and after (181/190) adapting granularity. Crucially, mapping guided active, sequential comprehension that prepared participants for modification (P3, P5, P7, P9, P17, P20): it was the most common action immediately before initiating a summary edit (60/79) or direct instruction (29/32). As P5 noted, it ``\textit{helped me not only understand the code but also plan how to change it}.''

\textbf{Mapping externalized the cognitive load of mentally tracing NL-code relationships and extended its value across the full modification lifecycle.} P2 noted that, while typical AI tools require ``\textit{the human brain to do the mapping},'' \system{} ``\textit{automatically did the mapping, which reduced the burden}.'' Developers also preferred faster localization by being able to ``\textit{just click on it, and it would take me to the corresponding chunk of code}.'' Backed by AST-based structural alignment, the mappings remained stable and accurate \added{throughout sessions}, even as the code evolved through manual edits. After \system{} generated a code change, developers' most common first action was to inspect the updated mapping (44/68), relying on it to validate the automated changes. As P12 noted, it could ``\textit{guide me to quickly verify if the summary is correct}.''\looseness=-1

\subsubsection{Viscosity, Visibility \& Consistency (RQ4)}
\label{sec:rq4}

\textbf{\system{}'s intent-driven workflow reduced the viscosity of intent specification by integrating the flexibility of direct instructions into summary editing.} 14/20 participants strongly agreed that expressing intent through high-level instructions was natural and efficient (Figure~\ref{fig:likert_non_comparison}), supported by significantly higher ratings for intent specification (Q6: $Mdn_N=6.0$, $Mdn_B=4.5$, $W=1.50$, $p=.002$). As shown in Figure~\ref{fig:interaction_transition}, 58/79 \texttt{Commit Modified Summary} actions followed directly from \texttt{Apply Instruction to Summary} without manual refinement, with P5 noting that it ``\textit{did a lot of the heavy lifting}.'' The intermediate summary diff further supported intent clarification: initial instructions were concise ($Mdn=12$ words), whereas the corresponding summary edits were significantly longer ($Mdn=18$ words, $W=556.0$, $p<.001$; see Figure~\ref{fig:word_diff_boxplot}). This expansion was not merely an implementation detail; it actively resolved ambiguity before code generation. For instance, when P1 vaguely instructed the system to find the \textit{best} NDCG score, the summary revised this to \textit{highest}, clarifying the metric's directionality. Similarly, P3 noted that the process turned a ``\textit{vague, conversational request}'' into a ``\textit{more specific summary}.'' Although the system occasionally misinterpreted subtle instructions, e.g., by hard-coding literal month abbreviations (``\textit{Jan, Feb}'') instead of inferring a general date-formatting rule (P3, P12), the intermediate layer made such failures visible and correctable before code generation.

\textbf{Bidirectional synchronization provided consistency and visibility that developers \added{valued highly}} (Figure~\ref{fig:likert_non_comparison}). \system{} received significantly higher ratings for validating modified code (Q8: $Mdn_N=6.0$, $Mdn_B=4.0$, $W=3.50$, $p=.001$) and supporting iterative refinement (Q9: $Mdn_N=6.0$, $Mdn_B=5.0$, $W=13.50$, $p=.004$). Participants compared the auto-updated diff to reviewing a GitHub commit for ``\textit{accidental changes}'' before merging (P6), and relied on it to ``\textit{see at a glance where the modification was}'' (P8) rather than ``\textit{re-reading everything}'' (P5).
\section{Discussion and Design Implications}

\subsection{Richer and Malleable NL Representations}

Our study showed that structured NL representations support comprehension: developers preferred bulleted summaries because they align with code's procedural logic (Section~\ref{sec:rq2}), consistent with Information Foraging Theory~\cite{pirolli1999information}, which posits that structured cues reduce the ``cost of access'' to relevant information. However, \system{} represents only a first step. As P2 observed, bulleted lists convey ``\textit{parallel and hierarchical relationships}'' but struggle to capture causality, while codebases are inherently non-linear, combining \textit{tree} structures (files, classes, and methods) with \textit{graph} structures (call dependencies and data flow). Future NL representations should therefore move beyond simple lists to incorporate richer forms that better reflect how developers navigate large codebases~\cite{ko2006exploratory, fleming2013information}.\looseness=-1

Beyond structure, these representations should also be malleable and personalized~\cite{richter2013malleable, cao2025generative, chen2026taskartisan}. From a cognitive perspective, \system{}'s six predefined variants already enable ``epistemic actions''~\cite{kirsh1994distinguishing}: developers reconfigure the representation to offload cognitive processing by adjusting information density to match their working memory capacity. Yet participants expressed a need for deeper customization; for example, P2 preferred function signatures over descriptive summaries at low granularity. This suggests that future systems should both implicitly learn developers' information needs across contexts (e.g., expertise, task complexity) and support proactive customization of representation types.

\subsection{Generalizing to Spec-Driven Development}
\label{sec:spec_driven}

While \system{} focuses on localized code modification, its design principles connect to the emerging paradigm of \textit{spec-driven development}, formalized in tools such as GitHub Spec Kit\footnote{\url{https://github.github.com/spec-kit/}} and Amazon Kiro\footnote{\url{https://aws.amazon.com/documentation-overview/kiro/}}, where lightweight NL documents (e.g., \texttt{design.md}, \texttt{tasks.md}) serve as the primary medium for human-AI interaction, guiding implementation and validating code against written constraints~\cite{tang2026programming}. Both workflows face a common challenge: maintaining alignment between an NL representation of intent, AI-generated code, and developer understanding~\cite{tang2026coding}. \system{}'s three design principles address this alignment problem: \textit{adaptive representation} supports navigating NL artifacts across levels of abstraction; \textit{interactive mapping} can link spec items to their corresponding implementations, making NL documents live and navigable rather than static; and \textit{intent-driven bidirectional synchronization} enables developers to modify NL to drive code changes while reflecting those changes back, creating a tighter feedback loop than current tools provide. More broadly, this points to a \textit{three-way alignment problem} common to both workflows: developers must verify not only that the code is correct, but also that the NL artifact captures their intent and that the AI's actions remain consistent with it. \system{}'s diff and validation workflow is designed to make this misalignment visible and resolvable.

\subsection{An Ecosystem for NL-Centric Programming}

Our findings point to two directions for extending \system{} into a broader programming ecosystem. First, NL representations should surface runtime errors: as P9 suggested, when code fails, the system should highlight the problematic logical segment in the summary, thereby integrating debugging into the comprehension-editing workflow~\cite{yang2026spark}. Second, the non-linear nature of AI-assisted exploration calls for lightweight personal history management beyond the current diff-based revert mechanism, including support for exploratory branching, path comparison, and selective merging. This may be particularly important for vibe coding~\cite{sarkar2025vibe}, where developers iteratively refine large-scale edits (P5, P8, P12, P17).
\section{Limitations and Future Work}

\paragraph{Implementation Constraints}
As a prototype, \system{} has several practical limitations. First, its interactive responsiveness is constrained by the latency of the underlying LLM. Although we applied optimizations such as request parallelization, feedback is not instantaneous and can disrupt the fluidity of the interaction loop. Second, our approach incurs substantial computational cost, which may be prohibitive for larger codebases. Future work should explore more efficient methods (e.g., pre-indexing) to establish coarse-grained alignment between code and summary before invoking a more expensive model for fine-grained mapping. \added{Third, summary sessions do not persist across IDE restarts; project-level persistence is left as future work.}

\system{} can also exhibit instability when regenerating summaries: as reflected in the expert ratings (Section~\ref{sec:expert_rating_results}), diff salience decreases at finer granularity, where great change volume and inconsistent phrasing obscure the most important edits. Future work should explore semantically controlled generation techniques, e.g., extending constrained decoding~\cite{willard2023efficient} from structurally simple settings to semantically richer representations. \added{Additionally, while AST anchor re-resolution degrades gracefully by design (Section~\ref{sec:structural_alignment_engine}), its robustness has yet to be systematically quantified.}

\paragraph{Threats to User Study Validity}

\added{Our within-subjects comparison used a feature-ablated baseline, which supports comparison against prior NL-mediated approaches but cannot isolate the contribution of individual mechanisms. We instead examined each feature through per-dimension ratings (Figure~\ref{fig:likert_non_comparison}), interview quotes, and transition analysis of interaction logs (Figure~\ref{fig:interaction_transition}), where actions map onto individual features. A full factorial ablation remains future work. We also did not compare against commercial AI coding tools (e.g., Cursor, Copilot), as differences in underlying models, mature UIs, and participants' prior proficiency would introduce confounds that obscure attribution. Our goal is to offer design principles that complement such tools rather than outperform them (Section~\ref{sec:spec_driven}), and we leave a direct comparison as future work.}

\added{Our findings on confidence and control are also subject to automation bias: a readable NL diff may increase perceived control while still concealing semantic errors or inaccurate NL-code mappings. Two observations partially mitigate this concern: task correctness with \system{} exceeded the baseline (95.0\% vs.\ 81.7\%), and the editable summary diff exposed misread instructions before code generation (Section~\ref{sec:rq4}). However, our logs do not capture the intent behind summary edits, so we could not quantify how often the diff misrepresented an instruction or how often developers caught such misreads; both questions warrant dedicated study.}

External validity is further limited by task scope and participant sample. Although participants found the tasks realistic, they were necessarily bounded by a lab session and could not capture the complexity of long-term software maintenance, reflecting the trade-off between experimental control and ecological validity in programmer studies~\cite{siegmund2015views, ko2015practical}. In addition, predefined modification goals bypassed the open-ended problem identification and diagnosis common in real-world development. Although our 20 participants varied in experience, the sample may still not represent the broader developer population.

Moreover, part of our analysis relies on self-reports, such as interviews and usability ratings. To mitigate this, we triangulated these findings with behavioral data \added{throughout Section~\ref{sec:user_study_results}: e.g., self-reported value of mapping by its 488 logged occurrences as the most frequent action}. As in many lab studies of interactive systems, participants' perceptions may also have been influenced by the novelty of the interaction design. Future work should therefore evaluate our approach through longitudinal, real-world deployments~\cite{yang2026lessons} with a larger and more diverse participant pool to examine how usage patterns evolve and whether novelty effects persist. Because \system{} is implemented as a VS Code extension, it is well-suited to such in-situ evaluation in realistic development settings.
\section{Conclusion}

We present \system{}, a system that operationalizes design principles for making static NL summaries usable as interactive representations for code modification. Grounded in the \textit{Cognitive Dimensions of Notations}, \system{} introduces three interaction mechanisms to reduce cognitive load and enhance developer control. Through a two-stage technical evaluation and a controlled user study with 20 developers, we show that the approach is both technically viable and improves the developer experience. Our work offers validated design principles for future NL programming tools and points toward richer representations of code.

\begin{acks}
This research was supported in part by a Google Cloud Research Credit Award, a Google Research Scholar Award, a gift from Adobe, an Edison Innovation Fellowship from the Notre Dame IDEA Center, and NSF grants CCF-2211428, CCF-2315887, CCF-2100035, CCF-2211429, and CCF-2442682. Any opinions, findings, or recommendations expressed here are those of the authors and do not necessarily reflect the views of the sponsors.
\end{acks}

\bibliographystyle{ACM-Reference-Format}
\bibliography{reference}

\appendix
\section{Data and Code Availability}

To support transparency and reproducibility, we have made our research artifacts publicly available in a replication package\footnote{\added{\url{https://github.com/ND-SaNDwichLAB/naturaledit-replication-package}}}.

This package contains: (1) the complete source code for \system{}\footnote{\added{Also available at \url{https://github.com/TTangNingzhi/NaturalEdit}.}} and \textit{Baseline}; (2) materials for the technical evaluation, including the benchmark evaluation and the expert rating; (3) materials for the user study, including study tasks and all anonymized raw data (interaction logs and questionnaire responses); and (4) analysis artifacts, including the scripts for data analysis and visualization. All figures, tables, and statistical results can be fully reproduced using the provided scripts. \added{The \system{} extension is available in the VS Code Marketplace.\footnote{\url{https://marketplace.visualstudio.com/items?itemName=sandwich-lab.naturaledit}}}


\section{Prompt Templates}
\label{sec:prompt_templates}

\begin{lstlisting}[style=prompt,caption={Prompts for Generating Multi-Faceted NL Representation}]
You are an expert code summarizer. For the following code, generate 6 summaries, one for each combination of detail level (low, medium, high) and structure (unstructured, i.e., paragraph, structured, i.e., bulleted):
- low_unstructured: One-sentence, low-detail, paragraph style.
- low_structured: 2-3 short bullet points, low-detail, as a single string. Each bullet must start with "•" and be separated by \n. Never return an array.
- medium_unstructured: 2-3 sentences, medium-detail, paragraph style.
- medium_structured: 3-5 bullet points, medium-detail, as a single string. Use "•" for first-level bullets, and ENCOURAGE the use of two-level bullets (use "◦" for the second level, and indent the second-level bullet with 2 spaces before the "◦") when logical groupings exist. Bullets must be separated by \n. Never return an array.
- high_unstructured: 3-4 sentences, high-detail, paragraph style.
- high_structured: 4-8 bullet points, high-detail, as a single string. Use "•" for first-level bullets, and ENCOURAGE the use of two-level bullets (use "◦" for the second level, and indent the second-level bullet with 2 spaces before the "◦") when logical groupings exist. Bullets must be separated by \n. Never return an array.

IMPORTANT:
- For medium_structured and high_structured, if there are logical groupings, you should use two-level bullets ("•" and "◦"). For the second-level bullet ("◦"), always indent with 2 spaces before the "◦".
- The file context below is provided ONLY for reference to help understand the code's environment.
- Your summary MUST focus ONLY on the specific code snippet provided.
- Return your response as a JSON object with keys: title, low_unstructured, low_structured, medium_unstructured, medium_structured, high_unstructured, high_structured.

File Context (for reference only):
${fileContext}

Code to summarize:
${code}
\end{lstlisting}

\begin{lstlisting}[style=prompt,caption={Prompts for Building Interactive Cross-Representation Mapping}]
You are an expert at code-to-summary mapping. Given the following code and summary, extract up to 10 key summary components (phrases or semantic units) from the summary.

IMPORTANT:
1. Each summaryComponent you extract MUST be a substring (exact part) of the summary text below.
2. Extract summaryComponents in the exact order they appear in the summary text.
3. Do NOT hallucinate or invent summary components that do not appear in the summary.

For each summaryComponent, extract one or more relevant code segments from the code that best match the meaning of the summary component.
- For each code segment, return both the code fragment (as a string) and its line number in the original code (1-based).
- Prefer to use a complete code statement (such as a full line, assignment, function definition, or block) as the code segment if it clearly represents the summary component's meaning.
- If a full statement is not appropriate or would be ambiguous, you should use a smaller, relevant fragment (such as a variable, function name, operator, or part of an expression).
- Only include enough code to make the mapping meaningful and unambiguous.
- If a code segment contains multiple lines, split them into separate objects in the codeSegments array.

Return as a JSON array of objects:
[
  { 
    "summaryComponent": "...", 
    "codeSegments": [
      { "code": "code fragment 1", "line": 12 },
      { "code": "code fragment 2", "line": 15 }
    ]
  },
  ...
]

Code (with line numbers for reference):
${codeWithLineNumbers}

Summary:
${summaryText}
\end{lstlisting}

\begin{lstlisting}[style=prompt,caption={Prompts for Applying Edit Instruction on Modifiable Summary}]
You are an expert at editing code summaries. In this scenario, a developer is using a summary-mediated approach to modify code:
1. Instead of directly editing the code, the developer modifies the summary to express their desired code behavior.
2. The modified summary will later be used to generate the actual code changes.
3. Your task is to integrate the developer's instruction into the summary, making it clear what the new code should do.

Given the following original summary and a direct instruction, update the summary to incorporate the developer's intent:
- The code context below is provided ONLY for reference to help understand the summary's environment.
- Preserve the parts of the original summary that are not affected by the instruction.
- Maintain the original summary format (sentence, bullet points, etc.).
- Make it easy to identify what changed by keeping unchanged parts exactly as they were.
- Integrate the instruction seamlessly into existing sentences or bullet points as much as possible.
- However, add new sentences or bullet points if the instruction cannot be naturally integrated into existing ones.
- The updated summary MUST clearly express what the new code should do, incorporating ALL information from the instruction.
- Output only the updated summary, nothing else.

Code Context (for reference only):
${originalCode}

Original summary:
${originalSummary}

Developer's instruction (integrate this intent FULLY into the updated summary):
${instruction}

Updated summary:
\end{lstlisting}

\begin{lstlisting}[style=prompt,caption={Prompts for Code Modification via Edited NL Summary}]
You are an expert code editor. Given the following original code and an updated summary (detail level: ${detailLevel}, structure: ${structuredType}), update the code to reflect the changes in the new summary.
- The file context below is provided ONLY for reference to help understand the code's environment, and your code changes MUST focus ONLY on the specific code snippet provided.
- Only change the code as needed to match the new summary, and keep the rest of the code unchanged.
- Preserve the leading whitespace (indentation) of each line from the original code in the updated code. For any modified or new lines, match the indentation style and level of the surrounding code.
- Pay close attention to the differences between the original summary and the edited summary, which reflects developer's intent of what the new code should be.
- Output only the updated code, nothing else.

File Context (for reference only):
${fileContext}

Original code:
${originalCode}

Original summary (detail level: ${detailLevel}, structure: ${structuredType}):
${originalSummary}

Updated summary (detail level: ${detailLevel}, structure: ${structuredType}):
${editedSummary}

Updated code:
\end{lstlisting}

\begin{lstlisting}[style=prompt,caption={Prompts for Generating NL Representation with Incremental Diffs}]
You are an expert code summarizer. Your task is to generate a new summary for the MODIFIED code below, using the original code and its previous summary as reference.

Instructions:
- Your new summary MUST focus on the code differences (addition, deletion) between the original and modified code and clearly reflect those changes, even if they are small, such as inline comments.
- Make the changed parts of the summary easy to identify (e.g., by being explicit about what changed, or by using wording that highlights the update). I mean, rather than describing the change itself (e.g., updated the function to ...), seamlessly integrate the changes into the new summary in one coherent, descriptive sentence.
- The new summary should be close to the old summary, only updating the parts that are affected by the code change:  If a part of the summary is still accurate for the new code, keep it unchanged; If a part of the summary is no longer accurate, change only that part to reflect the new code. Do not add unnecessary changes or rephrase unchanged parts.
- For all structured (bulleted) summaries, return as a single string. Each bullet must start with "•" and be separated by \\n. For medium_structured and high_structured, if there are logical groupings, you should use two-level bullets ("•" and "◦"). For the second-level bullet ("◦"), always indent with 2 spaces before the "◦". Never return an array.
- Return your response as a JSON object with keys: title, low_unstructured, low_structured, medium_unstructured, medium_structured, high_unstructured, high_structured.

File Context (for reference only):
${fileContext}

Original code:
${originalCode}

Old summary:
{
  "title": "${oldSummary.title}",
  "low_unstructured": "${oldSummary.low_unstructured}",
  "low_structured": "${oldSummary.low_structured}",
  "medium_unstructured": "${oldSummary.medium_unstructured}",
  "medium_structured": "${oldSummary.medium_structured}",
  "high_unstructured": "${oldSummary.high_unstructured}",
  "high_structured": "${oldSummary.high_structured}"
}

MODIFIED code:
${newCode}
\end{lstlisting}

\section{Technical Evaluation Materials}

\subsection{Benchmark Evaluation Results}
\label{sec:benchmark_evaluation_results_appendix}

The full results of the benchmark performance of the NL-mediated
modification (Section~\ref{sec:expert_rating_results}) are shown in Figure~\ref{fig:ratings_result}.

\begin{figure*}[htbp]
    \centerline{\includegraphics[width=0.8\textwidth]{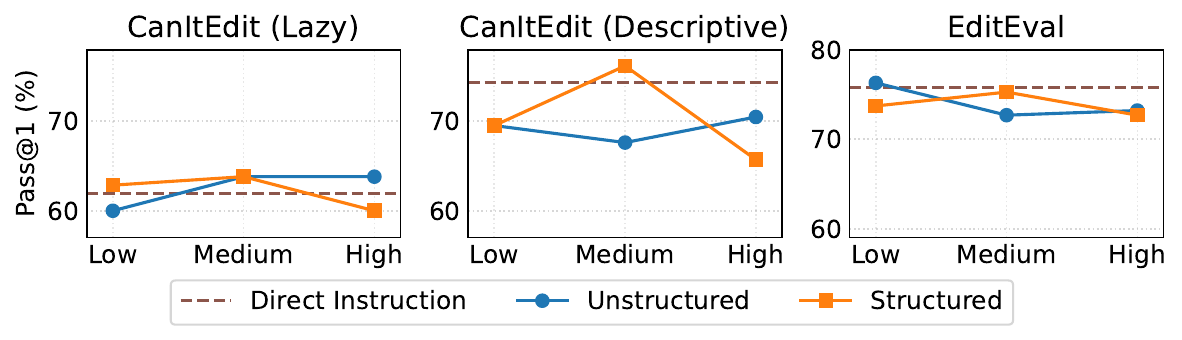}}
    \Description{A line chart with three panels showing Pass@1, the percentage of tasks passing all unit tests, on the vertical axis ranging from approximately 60 to 80 percent, across three natural language granularity levels (Low, Medium, High) on the horizontal axis. The three panels correspond to CanItEdit with Lazy instructions, CanItEdit with Descriptive instructions, and EditEval benchmarks. Each panel shows three lines: the Direct Instruction baseline as a dashed brown line, NL-mediated Unstructured as a blue line with circle markers, and NL-mediated Structured as a red line with square markers. Across all benchmarks, the natural-language-mediated lines track closely to the baseline with modest differences, confirming comparable technical performance.}
    \caption{Pass@1 of our NL-mediated workflows (blue and red lines) against the Direct Instruction baseline (dashed brown line), evaluated on CanItEdit~\cite{cassano2024can} (using both \textit{Lazy} and \textit{Descriptive} instructions) and EditEval~\cite{li2024instructcoder} benchmarks (Section~\ref{sec:expert_rating_results}).}
    \label{fig:datasets_result}
\end{figure*}

\subsection{Expert Rating Guide}
\label{sec:expert_rating_guide}

\subsubsection*{Overview}
This document provides the definitions and criteria for evaluating the quality of the intermediate artifacts generated by the \system{} system. The goal is to systematically assess the quality of the NL summaries, their mapping to the source code, and the diffs between summary versions.
All items will be rated on a 5-point scale, where the meaning of each point is defined for the specific dimension being evaluated. As a general guide:
\begin{itemize}
    \item \textbf{5: Excellent / Strongly Agree} - The artifact is of very high quality, with no significant issues.
    \item \textbf{4: Good / Agree}
    \item \textbf{3: Acceptable / Neutral} - The artifact is adequate for its purpose but has noticeable room for improvement.
    \item \textbf{2: Poor / Disagree}
    \item \textbf{1: Very Poor / Strongly Disagree} - The artifact is of low quality and has significant, potentially misleading, flaws.
\end{itemize}

\subsubsection*{Part A: Summary Quality}
\textit{(This is rated for each of the 36 buggy code summaries and the 36 ground-truth code summaries.)}

\textbf{A.1. Accuracy:}
Does the summary correctly describe the functionality and logic of the corresponding code?
\begin{itemize}
    \item \textbf{5 (Excellent):} Perfectly accurate. It correctly captures all key logic, operations, and outcomes without any errors or misinterpretations.
    \item \textbf{3 (Acceptable):} Mostly accurate. It captures the main purpose of the code but may contain minor omissions or slight inaccuracies that do not fundamentally mislead the reader.
    \item \textbf{1 (Very Poor):} Inaccurate. It contains significant factual errors, describes functionality that doesn't exist, or completely misses the core purpose of the code.
\end{itemize}

\textbf{A.2. Clarity:}
Is the summary written in clear, unambiguous, and easy-to-understand language for a developer?
\begin{itemize}
    \item \textbf{5 (Excellent):} Very clear, concise, and well-written. A developer could understand the code's function almost instantly.
    \item \textbf{3 (Acceptable):} Generally understandable, but the language might be slightly awkward, verbose, or contain jargon that could be simplified.
    \item \textbf{1 (Very Poor):} Unclear, confusing, grammatically incorrect, or so full of jargon that it is difficult to parse.
\end{itemize}

\subsubsection*{Part B: Segmentation \& Mapping Quality}
\textit{(This part has two components: ratings for each segmentation as a whole for 72 summaries, and ratings for each of the 478 individual mapping highlights.)}

\textbf{B.1. Segmentation Granularity:}
Is the code broken down into segments of an appropriate and useful size?
\begin{itemize}
    \item \textbf{5 (Excellent):} The granularity is perfect. Each segment is a meaningful, well-sized logical chunk.
    \item \textbf{3 (Acceptable):} The granularity is mostly fine, but some segments might be slightly too large (coarse-grained) or too small and fragmented.
    \item \textbf{1 (Very Poor):} The segmentation is not useful. It is either one giant, monolithic block or is excessively fragmented into meaningless single lines or tokens.
\end{itemize}

\textbf{B.2. Mapping Accuracy (Precision):}
Is this specific link that connects the summary segment to the code correct?
\begin{itemize}
    \item \textbf{5 (Correct):} The link is perfectly correct. The code highlight is directly relevant to the summary segment.
    \item \textbf{1 (Incorrect):} The link is wrong. The code highlight is irrelevant to the summary segment.
\end{itemize}

\textbf{B.3. Mapping Coverage (Recall):} 
Does this summary segment successfully link to all the relevant parts of the code that implement the concept it describes?
\begin{itemize}
    \item \textbf{5 (Complete Coverage):} Yes. The summary segment is linked to every piece of relevant code. No relevant code segments are missed.
    \item \textbf{3 (Minor Omissions):} Mostly. The main, most critical code blocks are linked, but a minor, less important part might be missing (e.g., a related variable declaration or a peripheral logging statement).
    \item \textbf{1 (Major Omissions):} No. Key code blocks that directly implement the summary segment's description are completely unlinked.
\end{itemize}

\subsubsection*{Part C: Summary Diff Quality}
\textit{(This is rated for each of the 36 diffs between the before-and-after summary pairs.)}

\textbf{C.1. Faithfulness:}
Does the change in the summary accurately represent the change that occurred in the code?
\begin{itemize}
    \item \textbf{5 (Excellent):} A perfect representation. The summary diff mirrors the semantic code change exactly.
    \item \textbf{3 (Acceptable):} Mostly faithful. It captures the main idea of the code change but might misrepresent a minor detail or nuance of the implementation.
    \item \textbf{1 (Very Poor):} Unfaithful. The diff is misleading, inaccurate, or completely unrelated to the actual change in the code.
\end{itemize}

\textbf{C.2. Completeness:}
Does the summary diff capture all of the important semantic changes from the code diff?
\begin{itemize}
    \item \textbf{5 (Excellent):} Fully complete. No important aspect of the code change is missing from the summary diff.
    \item \textbf{3 (Acceptable):} Mostly complete. It captures the primary functional change but might omit a secondary change (e.g., a change to a variable name or a log message that accompanies a logic change).
    \item \textbf{1 (Very Poor):} Incomplete. It completely misses one or more significant functional changes made in the code.
\end{itemize}

\textbf{C.3. Salience:}
Does the summary diff make the most important aspects of the change stand out clearly and effectively?
\begin{itemize}
    \item \textbf{5 (Excellent):} Highly salient. The change is presented clearly and concisely, making the core of the modification immediately obvious (e.g., a single, clear bullet point is added).
    \item \textbf{3 (Acceptable):} Moderately salient. The change is correctly represented, but it might be buried in other minor textual edits or phrased in a way that requires careful reading to understand the key point.
    \item \textbf{1 (Very Poor):} Not salient. The important change is obscured, minimized, or lost in a sea of trivial rewording, making it hard for a user to spot the key difference.
\end{itemize}

\subsection{Expert Rating Results}
\label{sec:expert_rating_results_appendix}

\begin{figure*}[htbp]
    \centering

    \begin{subfigure}[t]{0.8\textwidth}
        \centering
        \includegraphics[width=\linewidth]{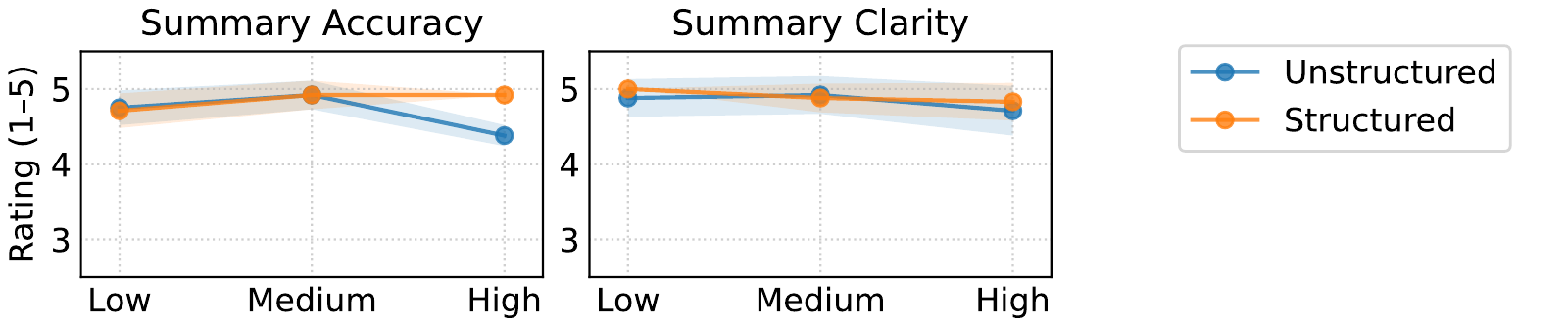}
        \caption{Summary quality across representation variants.}
        \label{fig:ratings_summary}
    \end{subfigure}

    \vspace{0.2em}

    \begin{subfigure}[t]{0.8\textwidth}
        \centering
        \includegraphics[width=\linewidth]{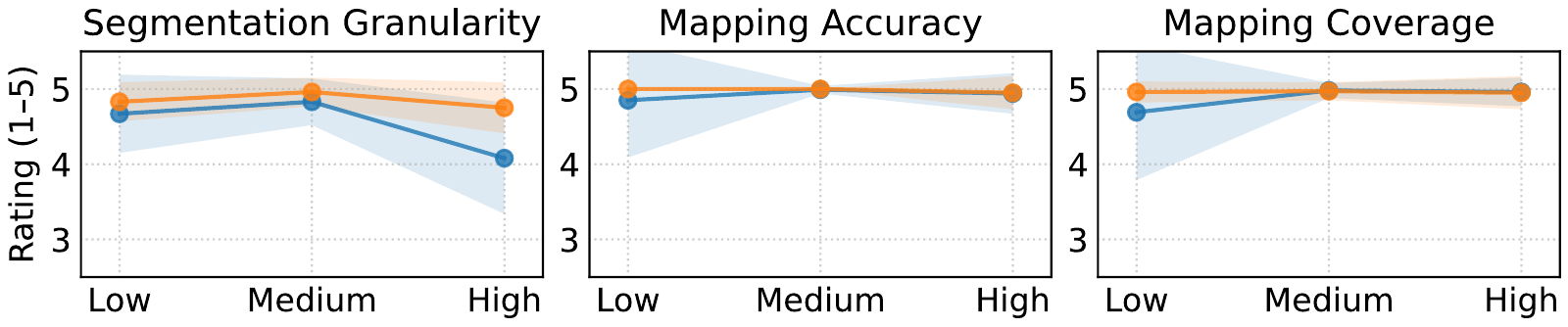}
        \caption{Mapping quality between NL summaries and code.}
        \label{fig:ratings_mapping}
    \end{subfigure}

    \vspace{0.2em}

    \begin{subfigure}[t]{0.8\textwidth}
        \centering
        \includegraphics[width=\linewidth]{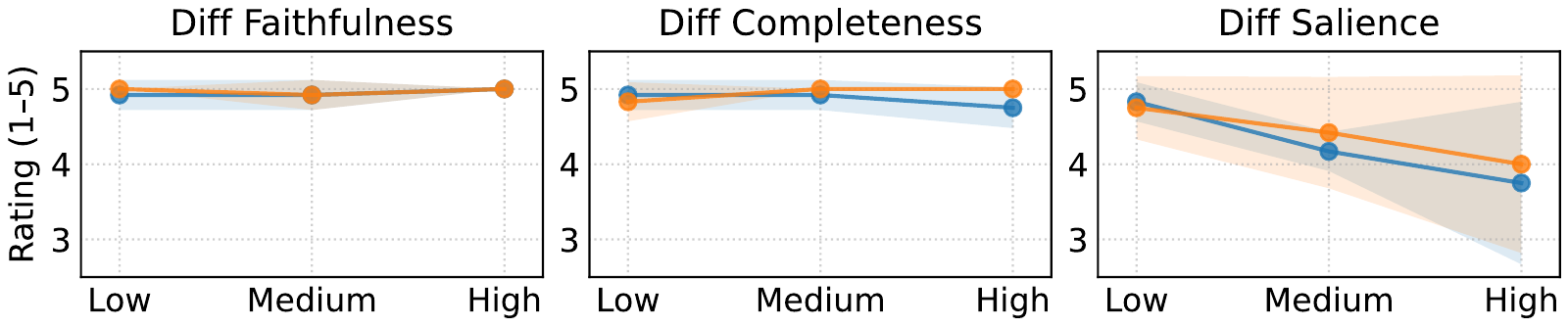}
        \caption{Diff quality for NL summary updates.}
        \label{fig:ratings_diff}
    \end{subfigure}

    \Description{Nine line charts arranged in three rows showing expert ratings on a 1 to 5 scale across granularity levels (Low, Medium, High) for Unstructured and Structured variants, with shaded regions representing standard deviations. Row (a) shows Summary Quality: both Accuracy and Clarity lines remain near 5.0 across all granularity levels with minimal variation. Row (b) shows Mapping Quality: Segmentation Granularity drops from near 5.0 at Low to approximately 4.0 at High with wider standard deviation bands, while Mapping Accuracy and Mapping Coverage remain near 5.0 throughout. Row (c) shows Diff Quality: Faithfulness and Completeness remain near 5.0 across all levels, while Salience declines notably from approximately 4.8 at Low to approximately 3.75 at High for the Unstructured variant, indicating that important changes become harder to identify at higher granularity.}
    \caption{Expert ratings of \system{} artifacts: points indicate mean ratings, and shaded regions represent standard deviations (Section~\ref{sec:expert_rating_results}).}
    \label{fig:ratings_result}
\end{figure*}

The full results of the expert evaluation of intermediate representations (Section~\ref{sec:expert_rating_results}) are shown in Figure~\ref{fig:ratings_result}.

\section{Controlled User Study Materials}

\subsection{Participants}
\label{sec:participants_appendix}

Table~\ref{tab:demographics} provides a detailed summary of participant demographics.

\begin{table}[htbp]
\centering
\caption{Summary of Participant Demographics.}
\Description{A table summarizing demographic information for 20 study participants labeled P1 through P20. Columns list ID, Gender, Age, Occupation, and Programming Experience. Ages range from 21 to 52 with a mean of 27.15. Occupations include PhD Student (11 participants), Software Engineer (4), R\&D Engineer (2), Graduate Student (1), and Undergraduate Student (2). Programming experience ranges from 3 to 40 years with a mean of 9.55 years. The gender distribution is 12 male and 8 female.}
\begin{tabular}{ccclc}
\toprule
ID & Gender & Age & Occupation & Experience \\
\midrule \midrule
P1 & Male & 23 & PhD Student & 5 years \\
P2 & Male & 25 & Software Engineer & 8 years \\
P3 & Male & 24 & Graduate Student & 6 years \\
P4 & Male & 25 & PhD Student & 6 years \\
P5 & Female & 27 & PhD Student & 11 years \\
P6 & Female & 52 & IT Professional & 30 years \\
P7 & Female & 22 & PhD Student & 5 years \\
P8 & Female & 22 & Software Engineer & 6 years \\
P9 & Male & 21 & Undergraduate Student & 3 years \\
P10 & Female & 21 & Undergraduate Student & 3 years \\
P11 & Male & 26 & R\&D Engineer & 7 years \\
P12 & Male & 24 & PhD Student & 8 years \\
P13 & Female & 28 & PhD Student & 7 years \\
P14 & Male & 23 & PhD Student & 6 years \\
P15 & Male & 48 & Software Engineer & 40 years \\
P16 & Male & 25 & PhD Student & 8 years \\
P17 & Female & 30 & Software Engineer & 10 years \\
P18 & Male & 30 & R\&D Engineer & 12 years \\
P19 & Female & 24 & PhD Student & 4 years \\
P20 & Male & 23 & PhD Student & 6 years \\

\bottomrule
\end{tabular}
\label{tab:demographics}
\end{table}

\subsection{Baseline}
\label{sec:appendix_baseline}

\begin{figure*}[htbp]
    \centerline{\includegraphics[width=0.9\textwidth]{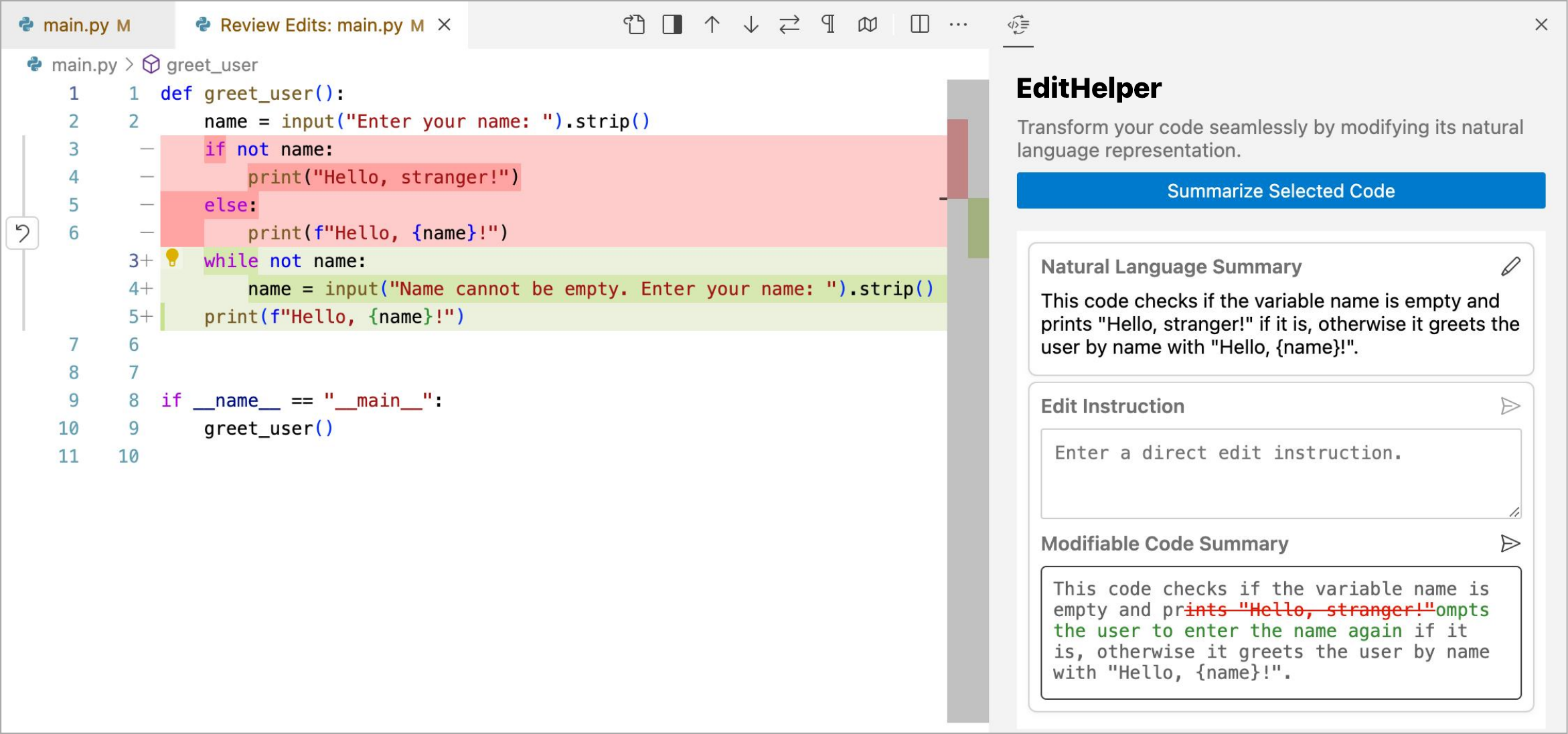}}
    \Description{A screenshot of the Baseline system named EditHelper, implemented as a VS Code extension with a side panel. The panel contains a ``Summarize Selected Code'' button, a static Natural Language Summary text block displaying a single paragraph, an Edit Instruction text box, and a Modifiable Code Summary editable text area. The editor on the left shows a Python file with a code diff. Unlike \system{}, the panel has no granularity slider, no structure toggle, and no interactive mapping controls.}
    \caption{A screenshot of the baseline system, implemented as an ablation of \system{}.}
    \label{fig:baseline_interface}
\end{figure*}

Figure~\ref{fig:baseline_interface} shows a screenshot of the baseline.

\subsection{Programming Tasks}
\label{sec:appendix_tasks}

\begin{figure*}[htbp]
    \centering
    \begin{subfigure}[t]{0.49\textwidth}
        \centering
        \includegraphics[width=\linewidth]{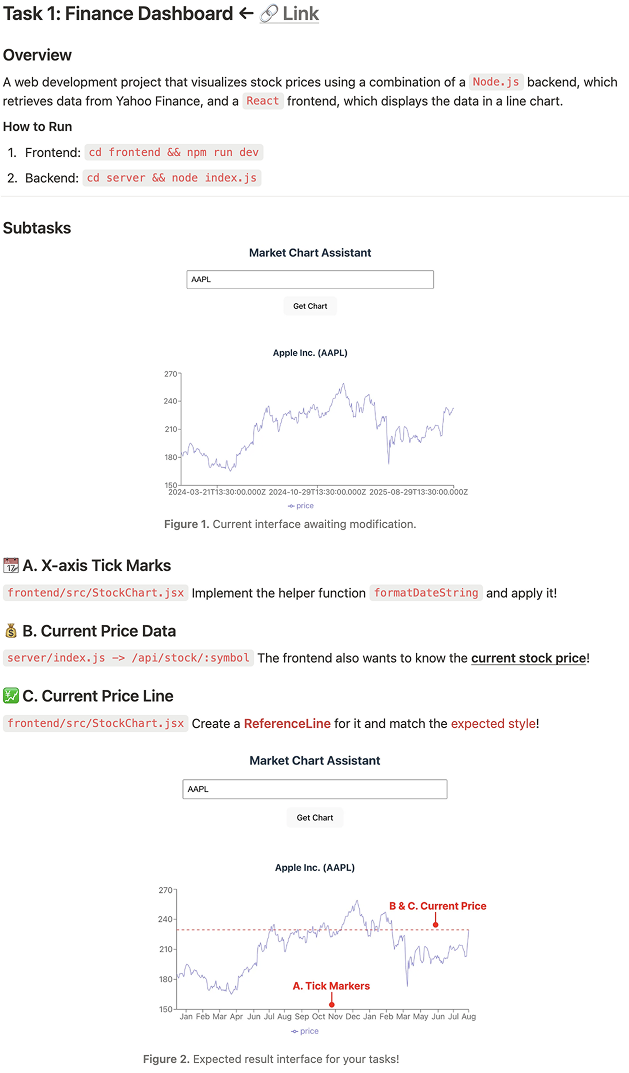}
        \caption{Task 1: Finance Dashboard}
        \label{fig:finance_dashboard}
    \end{subfigure}
    \hfill
    \begin{subfigure}[t]{0.49\textwidth}
        \centering
        \includegraphics[width=\linewidth]{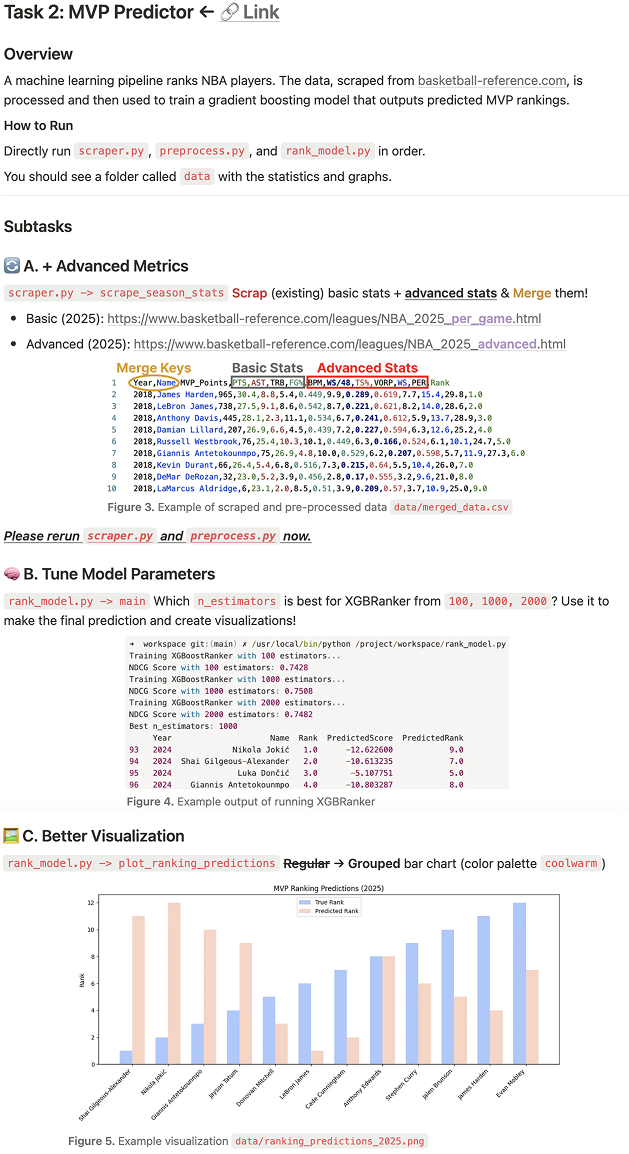}
        \caption{Task 2: MVP Predictor}
        \label{fig:mvp_predictor}
    \end{subfigure}
    \Description{Two screenshots of the study task descriptions side by side. Task 1 on the left shows the Finance Dashboard, a Node.js and React stock price visualization application retrieving data from Yahoo Finance. Three subtasks are labeled A, B, and C: Subtask A asks the participant to implement a \texttt{formatDateString} helper function for x-axis tick labels, Subtask B asks to add the current stock price to the backend API, and Subtask C asks to display the current price as a reference line on the chart. Annotated before-and-after chart screenshots illustrate the expected output. Task 2 on the right shows the MVP Predictor, a Python machine learning pipeline that scrapes NBA player data and predicts MVP rankings. Three subtasks are labeled A, B, and C: Subtask A asks to extend the scraper with advanced statistics by merging data from two URLs, Subtask B asks to find the best \texttt{n\_estimators} hyperparameter for an XGBRanker model among values of 100, 1000, and 2000, and Subtask C asks to change the visualization from a regular to a grouped bar chart using the \texttt{coolwarm} color palette. Example data tables and output visualizations are included for each subtask.}
    \caption{Screenshots of task descriptions used in the study.}
    \label{fig:tasks}
\end{figure*}

Figure~\ref{fig:tasks} presents screenshots of the task descriptions used in the study; full task code is provided in the replication package.

\subsection{Questionnaires}
\label{sec:questionnaires}

\subsubsection{NASA-TLX}
\textit{(1=Very low, 7=Very high) unless otherwise noted}

\begin{enumerate}[label=\arabic*.]
    \item \textbf{Mental Demand:} How mentally demanding was the task?
    \item \textbf{Physical Demand:} How physically demanding was the task?
    \item \textbf{Temporal Demand:} How hurried or rushed was the pace of the task?
    \item \textbf{Performance:} How successful were you in accomplishing what you were asked to do?
    \textit{(1=Perfect, 7=Failure)} \\ \textit{Note: Participants were informed that a lower score means perfect, while a higher score indicates failure.}
\end{enumerate}

\subsubsection{UMUX-LITE}
\textit{(1=Strongly disagree, 7=Strongly agree)}

\begin{enumerate}[label=\arabic*.]
\item This system's capabilities meet my requirements.
\item This system is easy to use.
\end{enumerate}

\subsubsection{Self-Defined Likert Scale Items.}
\textit{(1=Strongly disagree, 7=Strongly agree)}

\textbf{Post-Task Questions} \textit{(Asked immediately after each task, alongside the NASA-TLX)}
\begin{enumerate}[label=\arabic*.]
    \item I have a high level of understanding of the original code before it was modified.
    \item I have a high level of understanding of the code modifications made by the system.
\end{enumerate}

\textbf{Task Realism} \textit{(Asked once at the end of the entire session, the same applies below)}
\begin{enumerate}[label=\arabic*.]
    \item The study tasks felt as realistic as my daily programming.
\end{enumerate}

\textbf{Evaluation of both systems (\system{} and \textit{Baseline})}
\begin{enumerate}[label=\arabic*.]
    \item I could quickly learn how to use the system.
    \item I would use this system in my real development work if it were available.
    \item The system helps me comprehend the original code.
    \item The system supports me in specifying my intentions.
    \item The system helps me understand and validate the modified code.
    \item The system assists with the iterative refinement of code.
    \item The natural language representation is useful for achieving the task goal.
    \item I felt a good sense of control over the system's behavior.
    \item I am generally satisfied with the code modifications produced by the system.
\end{enumerate}

\textbf{Evaluation of \system{}'s specific features}
\begin{enumerate}[label=\arabic*.]
    \item The adaptive and multifaceted summaries helped in understanding the code at different levels.
    \item The interactive mapping between summary and code made their relationship explicit and easy to follow.
    \item By applying direct instructions to the summary, I was able to express my intentions flexibly and efficiently.
    \item The auto-updated summary with visual diffs helped me validate the changes in a consistent workflow.
\end{enumerate}

\subsection{Semi-Structured Interview Protocol}
\label{sec:interview}

\textit{The following questions served as a guide for our semi-structured interviews. This protocol was used flexibly; we often prioritized follow-up questions based on direct observations of a participant's behavior and their real-time commentary over strictly adhering to this script.}

\begin{itemize}[leftmargin=*, label={}]
    \item \textbf{Specific Observations}
    \begin{itemize}[leftmargin=*]
        \item ``I observed you did [action] when [situation]. Could you elaborate on your thought process there?''
    \end{itemize}

    \item \textbf{General Code Modification Workflow}
    \begin{itemize}[leftmargin=*]
        \item What are your main challenges when modifying existing, unfamiliar code (e.g., understanding its logic, validating changes)?
        \item What is your typical workflow for modifying existing code, and what are your main concerns during that process (e.g., introducing hidden bugs)?
    \end{itemize}

    \item \textbf{Feedback on \system{}'s Core Features}
    \begin{itemize}[leftmargin=*]
        \item Compare the experience of modifying code via its NL representation to your usual workflow of editing code directly.
        \item How did the different summary views (in terms of structure and granularity) affect your workflow? Did you have a preference?
        \item How useful was the interactive mapping between NL and code for understanding or modification?
        \item Did the automatically updated NL summary, with its visual diff, help you validate code changes?
    \end{itemize}

    \item \textbf{Perception of Output Quality and Control}
    \begin{itemize}[leftmargin=*]
        \item Did you notice any quality differences between the modified code and the generated NL summaries?
        \item Describe any moments where you felt more or less in control of the system's output.
    \end{itemize}
    
    \item \textbf{Contextual Factors}
    \begin{itemize}[leftmargin=*]
        \item How did your familiarity with the codebase affect your interaction with the system?
        \item Did the type of modification or the clarity of your goal influence how you used the system?
        \item When might NL-based features be most or least useful, depending on the code's complexity?
    \end{itemize}

    \item \textbf{Special Role} (For industry professionals)
    \begin{itemize}[leftmargin=*]
        \item How might this NL-based approach fit into your daily work, and what impact could it have?
    \end{itemize}

    \item \textbf{Concerns and Future Work}
    \begin{itemize}[leftmargin=*]
        \item What are the main drawbacks of \system{}, and what would prevent you from using it in your real work?
        \item Did anything about the study tasks feel artificial or unrealistic?
        \item How can this approach evolve to support large, multi-file projects?
    \end{itemize}
\end{itemize}

\begin{figure*}[htbp]
    \centerline{\includegraphics[width=\textwidth]{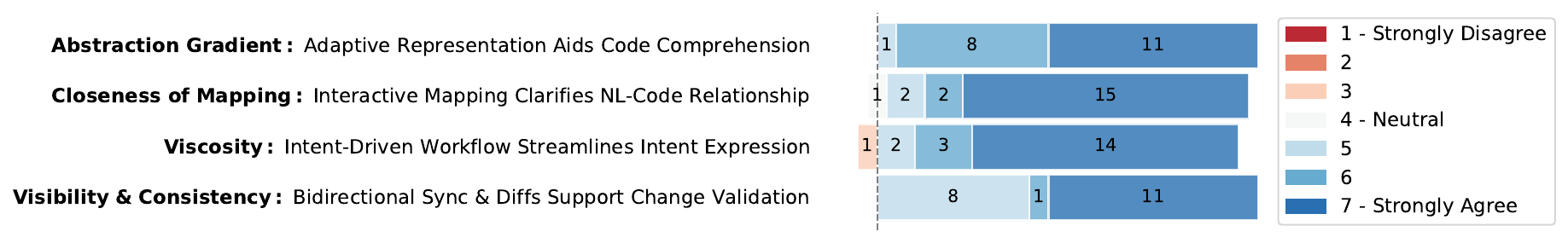}}
    \Description{A diverging stacked bar chart showing participant ratings from 1 (Strongly Disagree) to 7 (Strongly Agree) of \system{}'s four core interactive features, each associated with a cognitive dimension. Row 1, labeled Abstraction Gradient, asks whether adaptive representation aids code comprehension: 19 of 20 participants rated 6 or 7. Row 2, labeled Closeness of Mapping, asks whether interactive mapping clarifies the natural-language-to-code relationship: 15 of 20 participants rated 7. Row 3, labeled Viscosity, asks whether the intent-driven workflow streamlines intent expression: 14 of 20 participants rated 6 or 7. Row 4, labeled Visibility and Consistency, asks whether bidirectional sync and diffs support change validation: 11 of 20 participants rated 6 or 7, with 8 rating 5.}
    \caption{User ratings of \system{} features across four cognitive dimensions.}
    \label{fig:likert_non_comparison}
\end{figure*}

\begin{figure}[htbp]
    \centerline{\includegraphics[width=0.68\columnwidth]{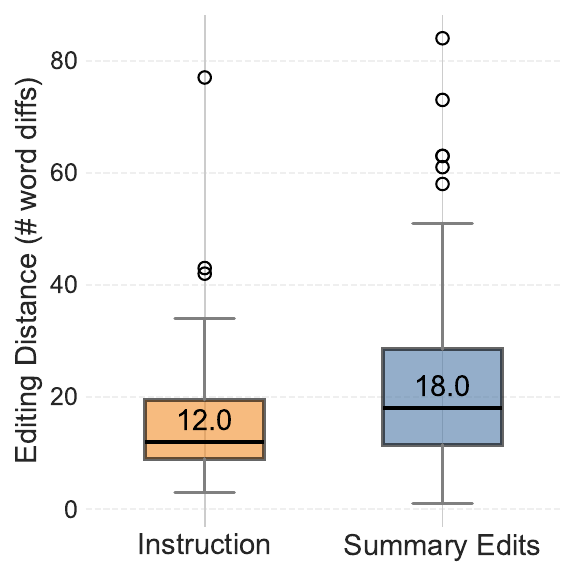}}
    \Description{A box plot comparing the editing distance measured in word-level differences using Gestalt pattern matching between participants' original instructions (left box) and the corresponding natural language summary edits generated by \system{} (right box). The vertical axis ranges from 0 to approximately 80 word differences. The instruction distribution has a median of 12.0 with a compact interquartile range. The summary edits distribution has a median of 18.0 with a wider interquartile range and several outliers above 60, indicating that the system consistently expands on brief user instructions.}
    \caption{The editing distance of the instruction and the corresponding summary edits generated by \system{}, across all summary granularities, measured as word-level differences using the Gestalt pattern matching~\cite{ratcliff1988pattern}. Median values are shown inside each box.}
    \label{fig:word_diff_boxplot}
\end{figure}

\subsection{Additional Study Results}
\label{sec:additional_study_results}

Figure~\ref{fig:likert_non_comparison} shows participants' ratings of \system{}'s four core interactive features along the corresponding cognitive dimensions.

Figure~\ref{fig:word_diff_boxplot} shows the distribution of word-level differences between participants' instructions and the corresponding summary edits generated by \system{}.

\subsection{Qualitative Codebook}
\label{sec:codebook}

\textit{This appendix presents the full codebook used for qualitative analysis. Codes are organized into thematic categories.}

\textbf{Comprehension Strategies for AI Code}
\begin{itemize}
  \item Different needs for understanding code semantics
  \item Avoid comprehension of code due to cognitive workload
  \item Preference for concise and clear summaries
  \item Reading the summary helps understanding
  \item Reading the summary involves a high workload
  \item Macro-level comprehension prioritized over line-level
  \item Top-down workflow: start with macro, then narrow to details
\end{itemize}

\textbf{Validation Strategies for AI Modifications}
\begin{itemize}
  \item Read the code to validate AI edits
  \item Run the code to validate AI edits
  \item Summarize edits with AI to assist validation
  \item Use AI to refine or fix edits directly
  \item Run tests to validate AI edits
\end{itemize}

\textbf{Direct Edit Instructions for AI Modification}
\begin{itemize}
  \item Direct instructions are natural and convenient
  \item Direct instructions are usually vague and casual
  \item Distrust in the accuracy of direct instructions
\end{itemize}

\textbf{Modifiable Code Summaries for AI Modification}
\begin{itemize}
  \item Manually editing summaries is a high workload
  \item Manual edits are more prone to error
  \item Summaries lack intuitive edit points
  \item Editing decisions depend on the simplicity of modifications
\end{itemize}

\textbf{Applying Edit Instruction to the Summary}
\begin{itemize}
  \item It makes editing intuitive and reduces workload
  \item Intermediate edited summaries pre-validate modifications
  \item Summaries help refine or clarify vague intent
  \item Summaries help understand LLM edits
  \item Summaries increase sense of control over edits
  \item Less trust in LLM's ability to focus on summary differences
  \item Preference for automation without manual confirmation
  \item Trade-off between controlling logic and trusting instructions
\end{itemize}

\textbf{Adjustable Granularity and Structure}
\begin{itemize}
  \item Different granularity adis different comprehension levels
  \item More granularity variations give more edit points
  \item Low granularity: easy to skim, useful for large-scale edits
  \item High granularity: detailed understanding, contains edit points, useful for modification
  \item Medium granularity balances detail and efficiency
  \item Granularity should align with code units (block, function)
  \item Structured summaries align with program logic 
  \item Structured summaries are easier to read
  \item Structured summaries support hierarchy and skim reading
  \item Structured summaries help locate edit points
  \item Structured summaries help with large-scale edits
  \item Paragraph summaries are cohesive and easy to read
  \item Paragraph is lengthy, jumbled together, and harder to parse
  \item Hierarchy and indentation keep summaries structured
  \item Summaries should be context-aware or customizable
  \item Content of different granularities needs careful design
  \item Order mismatches between summary and code structures
\end{itemize}

\textbf{Interactive Mapping Between Summaries and Code}
\begin{itemize}
  \item Mapping helps understand code more effectively
  \item Mapping provides evidence for edits and builds trust
  \item Mapping works as auto-generated comments
  \item Mapping helps locate modification points
  \item Mapping reduces cognitive workload
  \item Clicking mappings supports code navigation on demand
  \item Mapping may be better for structural blocks than lines
  \item Visualizing mappings in summary input area is helpful
  \item Structure and mappings benefit non-English speakers
  \item Need mappings between code diffs and summary diffs
\end{itemize}

\textbf{Auto-Updated New Summaries with Diffs}
\begin{itemize}
  \item Auto-updated summaries maintain workflow consistency
  \item New summaries with diffs help iterative modification
  \item Diffs support validation
  \item Low trust in diff quality
  \item Highlight colors may blend with mapping colors
\end{itemize}

\textbf{Factors Impacting System Usability}
\begin{itemize}
  \item Familiar code requires only direct instructions
  \item Unfamiliar code increases reliance on summaries
  \item Imperative favors instructions, declarative favors summaries
  \item Code needing more control benefits from summaries
  \item Reliance on AI may hinder coding skill learning
  \item Scoping AI modifications improves control and testability
\end{itemize}

\textbf{Future Improvements for \system{}}
\begin{itemize}
  \item Support for multiple files
  \item Large projects need structured and cross-file connections
  \item Multi-file requires coarser mapping
  \item Code diffs should allow accept/reject options
  \item Support search and locating modification points
  \item Auto-summarize opened file without manual selection
  \item Jupyter Notebook support
  \item Code edit history management
  \item Alternative representations (dependency graph, data flow)
  \item Add debugging feature
  \item Include multimodal support, e.g., images
  \item Runtime sandbox to catch errors early
\end{itemize}

\end{document}